\documentclass[aps,prb,superscriptaddress,reprint,showpacs,amsfonts,amsmath]{revtex4-1}
\usepackage[final]{graphicx}

\usepackage[usenames,dvipsnames]{xcolor}
\usepackage[normalem]{ulem}

\begin{document}
\title{Diffusion and Interdiffusion in Binary Metallic Melts}
\date\today
\def\dlr{\affiliation{Institut f\"ur Materialphysik im Weltraum,
  Deutsches Zentrum f\"ur Luft- und Raumfahrt (DLR), 51170 K\"oln,
  Germany}}
\def\zuko{\affiliation{Zukunftskolleg, Universit\"at Konstanz,
  78457 Konstanz, Germany}}
\def\udue{\affiliation{Soft Matter Laboratory, IPKM,
  Heinrich-Heine Universit\"at D\"usseldorf, Universit\"atsstr.~1,
  40225 D\"usseldorf, Germany}}
\def\uduedlr{\affiliation{Department of Physics,
  Heinrich-Heine Universit\"at D\"usseldorf,
  Universit\"atsstr.~1, 40225 D\"usseldorf, Germany}}
\author{P.~Kuhn}\dlr
\author{J.~Horbach}\udue
\author{F.~Kargl}\dlr
\author{A.~Meyer}\dlr
\author{Th.~Voigtmann}\dlr\uduedlr

\begin{abstract}
We discuss the dependence of self- and interdiffusion coefficients on
temperature and composition for two prototypical binary metallic melts,
Al-Ni and Zr-Ni, in molecular-dynamics (MD) computer simulations
and the mode-coupling theory of the glass transition (MCT).
Dynamical processes that are mainly entropic in origin
slow down mass transport (as expressed through
self diffusion) in the mixture as compared to the ideal-mixing contribution.
Interdiffusion of chemical species is a competition of slow kinetic modes
with a strong thermodynamic driving force
that is caused by non-entropic interactions.
The combination of both dynamic and thermodynamic effects
causes qualitative differences in the concentration dependence of
self-diffusion and interdiffusion coefficients.
At high temperatures, the thermodynamic
enhancement of interdiffusion prevails, while at low temperatures,
kinetic effects dominate the concentration dependence,
rationalized within MCT as the approach to its ideal-glass transition
temperature $T_c$.
The Darken equation relating self- and interdiffusion
qualitatively reproduces the concentration-dependence
in both Zr-Ni and Al-Ni, but quantitatively, the kinetic contributions to
interdiffusion can be slower than the lower bound suggested by the Darken
equation.
As temperature is decreased, the agreement with Darken's equation improves,
due to a strong coupling of all kinetic modes that is a generic feature
predicted by MCT.
\end{abstract}

\maketitle

\section{Introduction}

Mass transport processes in dense melts are governed by highly cooperative
phenomena \cite{Faupel.2003}. This makes physical modeling based on
a~priori information on their structural properties a challenge.
In the hydrodynamic regime of long times and large spatial scales,
these processes are characterized by a number
of transport coefficients. In the case of binary mixtures,
the self-diffusion coefficient $D^s_\alpha$ describes long-range
transport of a tracer particle (of a given species $\alpha$); a single
inter-diffusion coefficient $D_{cc}$ describes the decay of concentration
fluctuations on large scales; the shear viscosity $\eta$ reflects the
slow decay of microscopic stress fluctuations.
These coefficients enter the mesoscopic and macroscopic
modeling of materials; their values and parameter dependence remain
important assumptions in these approaches \cite{Kasperovich.2010}.
Understanding mixing effects, i.e., the dependence of these coefficients
on melt composition, is of utmost importance in designing materials, and
reveals crucial information on the underlying microscopic physical
mechanisms.

In this paper, we will investigate model binary metallic melts at various
temperatures and composition by molecular-dynamics (MD) computer simulation.
This technique allows to decompose the different contributions to
mass transport; \cite{Boon} in particular to the inter-diffusion coefficient
$D_{cc}=L\cdot\Phi$ that is composed of a purely kinetic (Onsager)
coefficient $L$, and a thermodynamic factor $\Phi$.
We will demonstrate that their composition dependence is qualitatively
different, so that interdiffusion processes are described by a competition
of two opposing forces, a dynamic and a thermodynamic one.
The dynamic contribution can be understood as the precursor of kinetic
arrest as described by the mode-coupling theory of the glass transition
(MCT) \cite{Goetze.2009},
and consequently depends sensitively on control parameters.

Reliable experimental data for the mass transport coefficients in metallic
melts are scarce, despite their importance.
Some self-diffusion coefficients are accessible in
quasi-elastic neutron scattering \cite{Boon,Meyer.2002,Brillo.2011}, but only for those
atomic species that have a strong incoherent contribution to the low-$q$
scattering signal.
Classical diffusion-couple experiments such as the long-capillary technique
measure the species' concentration profile some time after the relaxation
of a macroscopic concentration step, interpreted using the Fickian diffusion
law \cite{Shimoji.1986,Griesche.2004}. In-situ methods, only recently developed, such as X-ray radiography \cite{Zhang.2010} or neutron radiography \cite{Kargl.2011} are
required to ensure that the relevant transport mechanism that is probed
in the experiment is indeed diffusion. Still, these measurements are
plagued with experimental artefacts originating from free surfaces
and/or inhomogeneities \cite{Kargl.2013}.
In addition, buoyancy-driven convective flow
may necessitate experiments under micro-gravity conditions
\cite{Garandet.1995,*Barat.1996}.
This prohibits the systematic exploration of the melt's state space
encompassing temperature $T$ and number concentration of the species $x_\alpha$.

It is therefore tempting to establish relations between the different
mass transport coefficients, to be able to infer the remaining ones from
those that can be measured most reliably.
One famous example is the ``Stokes-Einstein''
relation (first derived by Sutherland \cite{Sutherland.1905})
connecting the self-diffusion coefficient of a macroscopic
Brownian particle to the shear viscosity of the suspending fluid,
$D^s\sim kT/\eta$.
Although valid only in the limit
of infinitely large tracers (on the scale of molecular interactions in the
fluid), the Stokes-Einstein relation has
been reported to hold reasonably well for the self-diffusion of the
fluid constituents themselves \cite{DeMichele.2001}.
Recent careful measurements of self-diffusion in liquid Zr-Ni suggest
that $D^s\cdot\eta\sim\text{const.}$ holds in a large temperature
range (without the $kT$ factor) \cite{Brillo.2011}.
This is interpreted as signalling
that the mechanisms governing self-diffusion and viscosity,
although at first sight rather different, are dominated by processes on
the same time scale. Within
MCT, these processes are identified with the slow
structural relaxation of density fluctuations.
Only for strongly supercooled glass formers, deviations are known in the form of a
``fractional'' Stokes-Einstein relation $D^s\sim\eta^{-\xi}$ (with $\xi<1$)
\cite{Fujara.1992,Ediger.2000}, which is a remarkable deviation from the hydrodynamic
origin of the original relation \cite{Zwanzig.1985}.
Despite the fortuitous character of the validity of the Stokes-Einstein
relation applied to self-diffusion, it remains a
useful concept for not-too-viscous melts, and forms the basis of
microscopic rheological measurements in many complex liquids
\cite{Puertas.2014}.

A similar remarkable link between collective transport
phenomena and tagged-particle motion is the so-called
Darken equation \cite{Darken.1948}. It relates collective interdiffusion to
the two self-diffusion coefficients. Labeling the species in a binary
mixture by $\alpha=\text{A}$ and $\text{B}$, one can write
\begin{multline}\label{darken}
  D_{cc}(x,T)=\left[
  x_\text{B}D_\text{A}^s(x,T)+x_\text{A}D_\text{B}^s(x,T)\right]
  \times \\ \times
  \Phi(x,T)\cdot S(x,T)\,,
\end{multline}
where we denote by $x_\alpha$ the number concentration of either species,
and use the symbol $x$ to denote a concentration dependence in cases where
the particle label is irrelevant due to $\sum_\alpha x_\alpha=x_\text{A}
+x_\text{B}=1$. The relation proposed by Darken on the basis of hydrodynamic
arguments, corresponds to setting $S=1$. In other words, the Onsager kinetic
coefficient
$L$ is approximated as the weighted average of the species' self-diffusion
coefficients, neglecting cross-correlation terms that arise in the collective
process \cite{Horbach.2007}.
The thermodynamic factor $\Phi$ is connected to
the second derivative of the Gibbs free energy $G$ w.r.t.\ the concentrations,
equivalently written as the zero-wavenumber limit of the inverse of the
concentration-fluctuation static structure factor $S_{cc}(q)$,
\begin{equation}\label{phi}
  \Phi(x,T)=\frac{x_\text{A}x_\text{B}}{kT}
  \frac{\partial^2G(x,T)}{\partial x_\text{A}\partial x_\text{B}}
  =\frac{x_\text{A}x_\text{B}}{S_{cc}(q\!=\!0;x,T)}\,,
\end{equation}
where $S_{cc}(q)=x_\text{B}^2S_{\text{A}\text{A}}(q)+x_\text{A}^2
S_{\text{B}\text{B}}(q)-2x_\text{A}x_\text{B}S_{\text{A}\text{B}}(q)$ in
terms of the (Ashcroft-Langreth \cite{Hansen})
partial static structure factors $S_{\alpha\beta}(q)$.
It has been noted already in the discussion of the original
contribution by Darken \cite{Darken.1948} that Eq.~\eqref{darken} (with
$S=1$)
involves implicit assumptions, no excess volume upon mixing being one of them.

The Darken equation contains the correct limits for
vanishing species concentration, i.e., $D_{cc}\to D^s_\text{minority}$
whenever $x_\alpha\to0,1$: the interdiffusion coefficient in the
presence of a dilute concentration of one of the species (in binary
mixtures) is given by the self-diffusion coefficient of this minority
species. Note that in this limit, also $\Phi\to1$:
For infinite dilution of one species, one expects this species to be
randomly distributed in the embedding host system formed by the majority
species. The Gibbs free energy then follows the ideal-mixing law,
$G\sim x\ln x+(1-x)\ln(1-x)$, leading to $\Phi=1$.
In a system that favors mixing, the intermediate-concentration minimum
will typically be deeper, so that $\Phi>1$ results for finite $x$.
This is the case for most dense metallic melts.

A notable feature of the Darken approximation is that it always estimates
the Onsager coefficient $L$ to be bounded by the two self-diffusion
coefficients, i.e., $\min(D^s_\text{A},D^s_\text{B})\le L
\le\max(D^s_\text{A},D^s_\text{B})$.
A ``correction factor'' $S$ is usually included in Eq.~\eqref{darken} to
account for non-ideal mixing effects and kinetic cross-correlations.
In the case of chemical diffusion
in crystals, it is called the Manning factor \cite{Manning.1961}.
In the liquid, it can be expressed in terms of the distinct parts of
the velocity autocorrelation functions \cite{Horbach.2007}, and can thus be
evaluated exactly from the MD simulation.
We will use $S$ to quantify the extent to which
the Darken approximation is valid or violated.

Mode-coupling theory establishes a structure--dynamics relationship
based on the equilibrium liquid-state static structure factors
\cite{Goetze.2009}. In principle,
given an effective interaction potential for some metallic melt, and following
some ad-hoc approximations, the theory
is able to calculate dynamical features in the viscous melt. It predicts
transport coefficients such as viscosity,
self-diffusivity, or interdiffusion, based on the assumption that the
slow relaxation typical for a viscous fluid is governed by a slow decay of
density fluctuations. MCT describes the effect that in dense liquids,
particles are transiently trapped in nearest-neighbor cages. Transport out
of these cages is possible by highly collective processes induced
by thermal fluctions, and these become ever more ineffective when the
density is increased or temperature is lowered. Eventually, this mechanism
described by MCT arrests at an ideal fluid-to-glass transition
at some critical temperature $T_c$. Approaching $T_c$ from above, diffusion
coefficients and inverse viscosity diminish much more quickly than expected
from an Arrhenius temperature dependence. For temperatures $T<T_c$
(where also strong deviations from the Stokes-Einstein relation are seen), other
transport mechanisms dominate that have been described as solid-like
\cite{Ehmler.1998,Zoellmer.2003,Faupel.2003};
these are not the aim of our discussion here.
They typically cause Arrhenius-like behavior with widely different activation
energies between the species \cite{Bartsch.2006,Bartsch.2010}.

Arguably the simplest model system for dense mixtures is the hard-sphere
system. Here one incorporates
only the core repulsion between atoms, modeled as an infinite energy barrier
that keeps particles separated by at least their hard-core diameter $d$.
All other interaction forces are set to zero. Consequently, temperature
does not change the thermodynamic state of the system, as there is no
intrinsic energy scale. The dynamics of the hard-sphere system becomes
slow if density is increased much like it slows down upon cooling in ordinary
melts. Hence, density and inverse temperature are often used interchangably
as a rule-of-thumb to discuss qualitative features of slow relaxation.
The appropriate measure for the number density of a hard-sphere system
is the dimensionless packing fraction, i.e., the fraction
of the sample volume filled by the hard spheres. This accounts for a trivial
change in number density when comparing systems with different particle
sizes.
The hard-sphere system has the additional advantage, that a
parameter-free, first-principles approximation for the static structure
factors exists in analytic form, namely the Percus-Yevick (PY) structure
factor \cite{Baxter.1970}. The predictions of MCT based on the PY approximation
for binary hard-sphere mixtures have been established in great detail
in the regime relevant for metallic melts
\cite{Barrat.1990,Goetze.2003,Foffi.2003,Foffi.2004}.

Based on the hard-sphere analogy, MCT suggests that the kinetic coefficients
at constant temperature display a minimum at intermediate $x$, while the
viscosity should display a similar maximum.
This arises, because for
binary mixtures of spheres with size ratio $\delta\gtrsim0.75$ and close
to unity, the increased disorder in the cages favors glass formation
at constant volume fraction \cite{Goetze.2003}.
For smaller size ratio, glass formation
becomes suppressed upon mixing, but this regime is not relevant for
metallic melts if one estimates effective hard-core sizes by, e.g.,
covalent radii.
Translating the hard-sphere result into a
change of the glass-transition temperature, the transition line
$T_c(x)$ can be expected to exhibit a maximum at intermediate $x$.
The maximum will occur in the vicinity of $x=1/2$, but the precise location
will depend on the size ratio \cite{Goetze.2003}.

In real metallic melts, the scenario is complicated by the fact that
the pure components will have different glass-transition temperatures
(would they not crystallize).
Hence, the expected minimum in kinetic coefficients
can be superimposed by an ``ideal-mixing'' trend that is either monotonically
increasing or decreasing.
In principle, MCT can predict the qualitative behavior of transport
coefficients based on experimental data for the equilibrium static
structure. However, this requires knowledge of all partial static
structure factors $S_{\alpha\beta}(q)$ over a reasonably large range
of wave numbers $q$. This is not easily achieved, in metallic melts
although in principle
possible, say, with neutron scattering where isotope substitution
allows to vary the relative scattering lengths of the constituents.
Three independent measurements then give all three elements of
the symmetric $2\times2$ matrix $\boldsymbol S(q)$ in the binary mixture,
onto which MCT calculations can be based.
This was so far carried out for a particular Zr-Ni composition
at a single temperature \cite{Voigtmann.2008b}.

The hard-sphere model serves as a reference to quantify purely entropic
effects of the dynamics. In realistic melts, non-entropic contributions
to the interactions result in stronger chemical short-range order.
The latter has been shown to change, e.g., the ratio of transport coefficients
\cite{Voigtmann.2008b}:
in a binary Zr-Ni melt, based on entropic contributions
one expects Ni diffusion to be faster than
Zr diffusion due to the smaller size of Ni atoms. Yet, taking into account
within MCT the strong Zr-Ni interactions that modify the partial static
structure factors, the theory predicts both diffusion coefficients to be
virtually identical, at least for the considered composition and temperature.
This qualitative change induced by non-entropic interactions asks for
a classification of various metallic melts according to the relation between
their transport coefficients (such as hard-sphere like or non-hard-sphere
like \cite{Meyer.2002}).

In this paper, we compare MD simulation predictions for the mass transport
coefficients of two exemplary binary metallic melts, Al-Ni and Zr-Ni, and their
composition dependence. These are model systems where experimental
investigations in the molten state are possible; first results are
available \cite{Das.2005,Horbach.2007,Voigtmann.2008b,Griesche.2009,Stueber.2010}, with
further measurements continuing to date.
We also use the static structure factors obtained by our MD simulations
as input to the MCT equations of motion, in order to compare the theory's
predictions for these specific melts with the dynamical results obtained
in the simulation. Doing so, we are able to identify the observed trends
of the mass transport coefficients upon mixing with some of MCT's generic
predictions.

The comparison of Al-Ni and Zr-Ni
allows to test which transport mechanisms are generic; both systems, while
sharing one atomic species, show very different glass-forming ability,
and very different phase diagrams. Dynamically, they represent
systems with a relatively weak effective size difference ($\delta
_\text{Al-Ni}\approx0.87$), respectively a relatively strong one
($\delta_\text{Zr-Ni}\approx0.78$).

The MD simulations are performed with recent embedded-atom potentials
that have been proven to be good models in terms of reproducing the melt's
characteristics with purely classical molecular dynamics. It should be
stressed that this modeling is intrinsically approximate. We therefore
do not claim our results to be quantitative predictions for \emph{real}
Al-Ni or Zi-Ni melts, but rather for model systems that allow to understand
the relevant mass transport mechanisms in such melts qualitatively and
semi-quantitatively.

This paper is organized as follows: we will recollect the details of
the MD simulations and mode-coupling theory analysis in Sec.~\ref{methods}.
Results are discussed in Sec.~\ref{results} first for the Al-Ni MD
simulations, then for the Zr-Ni simulations, and finally for the
mode-coupling calculations based on the MD-simulated static structure
factors for both systems. Finally, Sec.~\ref{conclusions} contains some
conclusions.

\section{Methods}\label{methods}

\subsection{MD Simulation}

Molecular-dynamics (MD) simulations were carried out using effective
interaction potentials of the embedded atom type (EAM).
One writes the total potential energy of a binary system as
\begin{equation}
  U=\frac12\sum_{i,j\neq i}V_{\alpha(i)\beta(j)}(r_{ij})
  +\sum_i F_{\alpha(i)}(\bar\rho_i)\,,
\end{equation}
where $\alpha(i)$ denotes the species to which particle number $i$ belongs.
$V_{\alpha\beta}(r)$ is the pair interaction potential between species
$\alpha$ and $\beta$, assumed to depend on the distance between the particles,
$r_{ij}=|\vec r_i-\vec r_j|$, only.
The embedding energy $F_{\alpha(i)}$ of species $\alpha$ at site $\vec r_i$
is given by the electron density $\bar\rho_i$ from all other atoms.
The latter is written as
\begin{equation}
  \bar\rho_i=\sum_{j\neq i}\rho_{\alpha(j)}(r_{ij})\,,
\end{equation}
the sum over the electron densities at site $\vec r_i$ due to a particle of
species $\alpha$ at site $\vec r_j$.
The exact functional forms
of the pair potential $V_{\alpha\beta}(r)$, of the electron
densities $\rho_\alpha(r)$, and of the embedding energies $F_\alpha(\rho)$
are empirical choices usually based on ab~initio simulations and
crystallographic data.

For Al-Ni,
a potential proposed by Mishin \textit{et~al.} \cite{Mishin.2002} was used,
as in earlier work of some of the present authors
\cite{Das.2005,Horbach.2007,Kerrache.2008,Das.2008}. We only summarize
the main aspects of the simulation here, and refer to these publications
for details.
MD simulations were performed by equilibrating the starting configurations
in the $NVT$ ensemble (with $N=1500$ particles),
with a temperature-dependent volume corresponding
to the average at a given pressure $p$, as determined from $NpT$ Monte Carlo
simulations. After that, microcanonical MD simulations were performed
to extract the dynamical quantities. Simulation runs covered more than
$10^5$ time steps using the velocity-Verlet integrator with time steps
varying between $1\,\text{fs}$ ($T\ge1500\,\text{K}$) and $2.5\,\text{fs}$
(lower $T$). At each temperature, $T=1795\,\text{K}$, $1500\,\text{K}$,
$1250\,\text{K}$, $1000\,\text{K}$, and $900\,\text{K}$, eight runs with
independent initial configurations were averaged over to improve
statistics. Compositions were chosen as $x_\text{Al}=0.1$, $0.25$, $0.4$,
$0.5$, $0.7$, $0.8$, and $0.9$ at the highest temperature. At lower
temperatures data were only obtained
for $x_\text{Al}=0.1$, $0.25$, $0.4$, and $0.8$.
The self-diffusion and inter-diffusion coefficients are
obtained via the appropriate Einstein relations from the long-time behavior
of the mean-squared displacement and the corresponding interdiffusion
quantity, see Ref.~\onlinecite{Horbach.2007} for details.

The potential for Zr-Ni was originally developed by Kumagai \textit{et~al.}
\cite{Kumagai.2007}, employing for fitting the structural information
of amorphous $\text{Zr}_\text{70}\text{Ni}_\text{30}$, and mainly
lattice parameters of the crystalline state. To improve the potential
for the molten state we are interested in, we have adjusted the
parametrization of Ref.~\onlinecite{Kumagai.2007} to reproduce quantitatively
experimental Ni-diffusion data \cite{Kordel.2011}. As further information,
static structure factors of $\text{Zr}_\text{64}\text{Ni}_\text{36}$
were taken into account \cite{Voigtmann.2008b}; here, perfect agreement
could not be enforced since a strong pre-peak emerges in the static structure
factor of the melt that is much weaker in the MD simulation.
Compared to Ref.~\onlinecite{Kumagai.2007}, the
potential energy was scaled by an overall factor $\alpha=1.86$, and
the parameter $f_{e,\text{Zr}\text{Ni}}$, describing the electron-density
ratio of atomic species Zr to Ni, was changed from the value $f_e=0.215$
determinde by Kumagai \textit{et~al.}, to $f_e=0.79$. For details on
the parameters entering the Zr-Ni potential, we refer to
Ref.~\onlinecite{Kumagai.2007}.

The glass-transition dynamics in Zr-Ni melts has been studied previously
by MD simulation combined with an analysis in terms of MCT by Teichler
and coworkers \cite{Teichler.1996,Mutiara.2001,Ladadwa.2006,Teichler.2011}.
The effective MD potential employed in these and related studies
differs from the one we use here.
We will remark on relevant differences below, where appropriate.

For the Zr-Ni system, simulations were run at
$x_\text{Zr}=0.36$, $0.5$, and $0.64$
over a temperature window covering the onset of slow dynamics and about
two orders of magnitude of slowing down. At $T=1400\,\text{K}$, additional
compositions $x_\text{Zr}=0.2$, $0.43$, $0.57$, and $0.8$
were considered. To discuss isothermic concentration
dependences, the results for the different compositions obtained from runs at
different temperatures $T<1400\,\text{K}$
were interpolated linearly to constant $T$.

\subsection{Mode-Coupling Theory}

A recent monograph \cite{Goetze.2009} describes the mode-coupling theory
of the glass transition and its application to experiments on molecular and
hard-sphere-like colloidal glass formers in detail. MCT calculations based
on MD-simulated partial static structure factors have been performed
before (see Ref.~\onlinecite{Sciortino.2001} for an early example). Details
for the present application follow those presented earlier for the
$\text{Al}_\text{80}\text{Ni}_\text{20}$ simulation \cite{Das.2008}.

MCT assumes that the slow dynamics of a dense melt is governed by the
relaxation of density fluctuations. The statistics of the latter are
encoded in the partial dynamic structure factors, or collective intermediate
scattering functions to wave vector $\vec q$,
\begin{equation}
  S_{\alpha\beta}(q,t)=\frac1N\sum_{k_\alpha=1}^{N_\alpha}
  \sum_{l_\beta=1}^{N_\beta}\langle\exp[i\vec q\cdot
  [\vec r_{k_\alpha}(t)-\vec r_{l_\beta}(0)]\rangle\,,
\end{equation}
where $\vec r_{k_\alpha}(t)$ marks the position of the particle labeled
$k_\alpha$ at time $t$.
At $t=0$ these functions yield the partial static structure factors
$S_{\alpha\beta}(q)$.
For an isotropic, translationally invariant equilibrium system,
the dynamic structure factors depend on the wave vectors only through
its magnitude $q=|\vec q|$, and on the time difference between the
two density fluctuations. Following a projection operator scheme,
an equation of motion is derived for the matrix
$\boldsymbol S(q,t)$ (see Ref.~\onlinecite{Goetze.2003}),
\begin{multline}
  \boldsymbol J^{-1}(q)\partial_t^2\boldsymbol S(q,t)
  +\boldsymbol S^{-1}(q)\cdot\boldsymbol S(q,t)
  \\
  +\int_0^t\boldsymbol M(q,t-t')\cdot\partial_{t'}
  \boldsymbol S(q,t')\,dt'=\boldsymbol 0\,.
\end{multline}
The matrix
$J_{\alpha\beta}(q)=q^2v_{\text{th},\alpha}^2\delta_{\alpha\beta}
=q^2k_BT/m_\alpha\delta_{\alpha\beta}$ sets
the thermal velocities governing the short-time relaxation.
The long-time relaxation is dominated by retarded-friction effects
that arise from slow collective dynamics. In the MCT
approximation, they are captured through a memory kernel that
is a nonlinear functional of the density correlation functions,
\begin{multline}
  M_{\alpha\beta}(q,t)=\frac1{2q^2}\frac{n}{x_\alpha x_\beta}
  \int\frac{d^3k}{(2\pi)^3}
  \sum_{\alpha'\beta'\alpha''\beta''}
  V_{\alpha\alpha'\alpha''}(\vec q,\vec k)
  \times\\
  V_{\beta\beta'\beta''}(\vec q,\vec k)
  S_{\alpha'\beta'}(k,t)S_{\alpha''\beta''}(p,t)
\end{multline}
with $p=|\vec q-\vec k|$ and $n$ the total number density. For the
coupling vertices we get
$V_{\alpha\alpha'\alpha''}(\vec q,\vec k)=(\vec q\cdot\vec k/q)
c_{\alpha\alpha'}(k)\delta_{\alpha\alpha''}
+(\vec q\cdot\vec p/q)c_{\alpha\alpha''}(p)\delta_{\alpha\alpha'}$
after neglecting a part that depends on the (unknown) static triplet
correlation function. In this approximation, the static-structure
factor matrix alone, through the related Ornstein-Zernike direct
correlation function $c_{\alpha\beta}(q)=(1/n)(\delta_{\alpha\beta}/x_\alpha
-(S^{-1})_{\alpha\beta})$, is sufficient to fully
determine the MCT equations of motion.

To calculate self-diffusion coefficients, one needs to characterize
the tracer-particle dynamics in the dense melt. This is achieved using
the self-part of the intermediate scattering function,
\begin{equation}
  \phi^s_\alpha(q,t)=\frac1{N_\alpha}\sum_{k=1}^{N_\alpha}
  \langle\exp[i\vec q\cdot[\vec r_k(t)-\vec r_k(0)]\rangle\,.
\end{equation}
The equation of motion for this tagged-particle density correlation function
is similar to its collective counterpart,
\begin{multline}\label{eomphis}
  \frac{1}{q^2v_{\text{th},\alpha}^2}\partial_t^2\phi^s_\alpha(q,t)
  +\phi^s_\alpha(q,t)\\
  +\int_0^tM^s_\alpha(q,t-t')\partial_{t'}
  \phi^s_\alpha(q,t')\,dt'=0\,.
\end{multline}
Here, the retarded-friction memory kernel is given, in the MCT approximation,
by a combination of the host-system density fluctuations and those of the
tagged particle, hence
\begin{multline}
  M^s_\alpha(q,t)=\frac n{q^2}\int\frac{d^3k}{(2\pi)^3}
  \sum_{\alpha'\beta'}
  (\vec q\cdot\vec k/q)^2c_{\alpha\alpha'}(k)c_{\alpha\beta'}(k)
  \times\\
  S_{\alpha'\beta'}(k,t)\phi^s_\alpha(p,t)\,.
\end{multline}
In the $q\to0$ limit, the tagged-particle density correlation function
is related to the mean-squared displacement (MSD) $\delta r^2_\alpha(t)$ by
$\phi^s_\alpha(q,t)=1-q^2\delta r^2_\alpha(t)/6+{\mathcal O}(q^4)$.
One readily verifies that for the MSD, an equation similar to
Eq.~\eqref{eomphis} holds,
\begin{equation}\label{mctmsd}
  \partial_t\delta r^2_\alpha(t)+v_{\text{th},\alpha}^2\int_0^t
  m^s_\alpha(t-t')\delta r^2_\alpha(t')\,dt'
  =6v_{\text{th},s}^2t\,,
\end{equation}
with
$m^s_\alpha(t)=\lim_{q\to0}q^2M^s_\alpha(q,t)$.
The diffusion coefficient follows from $\delta r^2_\alpha(t\to\infty)
\sim6D_\alpha t$, and is explicitly calculated as
\begin{equation}\label{diffmct}
  D_\alpha=\frac{1}{\int_0^\infty m^s_\alpha(t)\,dt}\,.
\end{equation}

In a multi-component system, the memory kernels $M_{\alpha\beta}(q,t)$
display a $1/q^2$ divergence for $q\to0$, which is connected with the
fact that particle numbers are conserved, but momentum of the individual
species is not conserved. As a result, one obtains the hydrodynamic modes
that are connected to interdiffusion.
Writing $N_{\alpha\beta}(t)=\lim_{q\to0}
q^2M_{\alpha\beta}(q,t)$, we obtain
\begin{equation}
  N_{\alpha\beta}(t)=\frac1{x_\alpha x_\beta}\int dk\,
  \sum_{\alpha'\beta'}
  V^0_{\alpha\beta\alpha'\beta'}(k)\,\Delta S_{\alpha\beta\alpha'\beta'}(k,t)
\end{equation}
with $\Delta S_{\alpha\beta\alpha'\beta'}=S_{\alpha\beta}
S_{\alpha'\beta'}-S_{\alpha'\beta}S_{\alpha\beta'}$ and
$V^0_{\alpha\beta\alpha'\beta'}=(n/(6\pi^2))k^4c_{\alpha\alpha'}(k)
c_{\beta\beta'}(k)$. Specifically for binary mixtures,
\begin{equation}\label{nab}
  N_{\alpha\beta}(t)=\frac{(-1)^{\alpha+\beta}}{x_\alpha x_\beta}
  \int dk\,V^0_{\text{A}\text{A}\text{B}\text{B}}(k)
  \det\boldsymbol S(k,t)\,.
\end{equation}
This encodes the symmetries of the interdiffusion process and highlights
that in a binary mixture, only one independent interdiffusion mode appears
in the low-$q$ dynamics of density fluctuations.

The interdiffusion coefficient
$D_{cc}$ is obtained from the concentration fluctuations
$\varrho_c=x_\text{B}\varrho_\text{A}-x_\text{A}\varrho_\text{B}$,
as
\begin{equation}\label{mctdcc}
  D_{cc}=\frac1{S_{cc}(0)\int_0^\infty m_{cc}(t)\,dt}\,,
\end{equation}
where
$m_{cc}(t)=N_{\text{A}\text{A}}(t)/x_\text{B}^2$.
For $x\to0$ or $x\to1$, the MCT expression for $D_{cc}$, Eq.~\eqref{mctdcc}
reduces to the one for the self-diffusion coefficient of the minority
species, Eq.~\eqref{diffmct}: if, say, $x_\text{B}\to0$, there holds
$S_{\text{B}\text{B}}(k,t)\sim x_\text{B}\phi^s_\text{B}(k,t)$ and
$\det\boldsymbol S(k,t)/x_\text{B}
\to S_{\text{A}\text{A}}(k,t)\phi^s_\text{B}(k,t)$,
so that the memory kernel $x_\text{B}N_{\text{B}\text{B}}(t)$
in Eq.~\eqref{nab} becomes identical to the memory kernel
$m^s_\text{B}(t)$ in Eq.~\eqref{mctmsd}.

At the glass transition, MCT predicts that all memory kernels become
non-decaying functions of time if they couple
sufficiently strongly to collective density fluctuations. Hence, the
integral in the denominators of Eq.~\eqref{diffmct} and \eqref{mctdcc} diverge,
so that $D_{cc},D_\alpha\to0$ as one approaches the glass transition from the
liquid side. In the present context, we do not discuss the possibility of
weak coupling of one species, which may cause the corresponding
diffusion coefficient to be non-zero in the glass
\cite{Voigtmann.2009,Voigtmann.2011b}.
This possibility makes clear that self- and interdiffusion are
governed by tagged-particle respectively collective density fluctuations
that are in principle different aspects of the dynamics, and might in fact
decouple. However, in typical glass-forming melts above and close to $T_c$,
the coupling between
these relaxation modes is so strong, that approximations connecting one to
the other may hold rather well.


In writing the MCT equations of motion as above, we have tacitly neglected
short-time contributions to the memory kernels that are expected to be
subdominant close to the MCT glass transition. These include a damping term
that is responsible for the behavior of the diffusion coefficients in the
less dense liquid, and is the object of classical liquid-state theory
\cite{Beijeren.1973,*Beijeren.1973a,Foffi.2003}.
It may in fact obey opposite mixing trends as the MCT contribution.
In the following, we focus on the low-temperature
dynamics, so that this omission will not change the results qualitatively.

\section{Results}\label{results}

\subsection{Al-Ni Simulations}

\begin{figure}
\includegraphics[width=.9\linewidth]{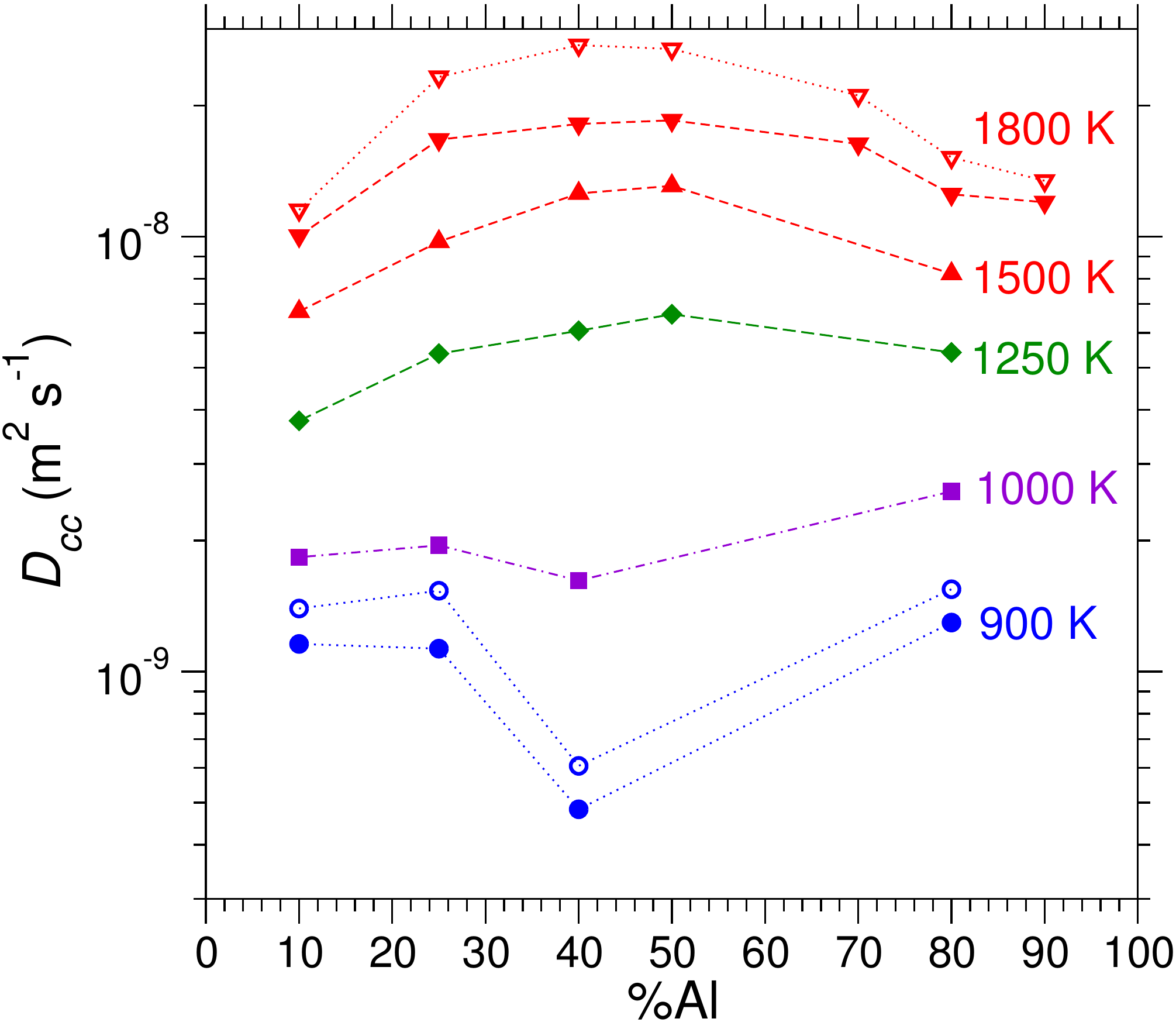}
\caption{\label{alnidx}
  Interdiffusion coefficient $D_{cc}$ in a model Al-Ni system, from
  molecular-dynamics computer simulations, as a function of Al concentration
  at various fixed temperatures (filled symbols, as labeled).
  Open symbols show $D_{cc}$ estimated from the Darken relation,
  Eq.~\eqref{darken} (without a correction, $S=1$) for $T=1800\,\text{K}$
  and $T=900\,\text{K}$.
}
\end{figure}

Figure~\ref{alnidx} shows MD simulation results for the interdiffusion
coefficient in $\text{Al}_x\text{Ni}_{100-x}$ alloys as a function of
the number concentration $x$ of $\text{Al}$ atoms. Different curves
correspond to changing concentration at fixed temperature.
The data extend those discussed in Ref.~\onlinecite{Griesche.2009}, where
only a single temperature was studied.
As the temperature is lowered, interdiffusion becomes slower at any
concentration. This is the typical trend for all kinetic transport coefficients,
and has been discussed for the present simulation before \cite{Horbach.2007}.

At high temperatures, $T=1500\,\text{K}$, say, one observes a maximum
in the interdiffusion coefficient $D_{cc}$ as a function of composition:
for Al concentrations around
$x\approx0.4$ to $x\approx0.5$, interdiffusion is enhanced by roughly a
factor of $2$ with respect to the systems containing only a small
concentration of either $\text{Al}$ or $\text{Ni}$.

This concentration dependence of $D_{cc}$ changes qualitatively as a function of
temperature. Approaching
$T\approx1000\,\text{K}$, the maximum observed at higher temperatures
vanishes, and at $T=900\,\text{K}$
instead of a maximum, a minimum is observed at $x\approx0.4$. At this
temperature, interdiffusion is more than a factor $2$ \emph{slower} in the
$\text{Al}_\text{40}\text{Ni}_\text{60}$ system than in any of the
almost pure systems.
This change from a maximum in $D_{cc}$ due to mixing, to a minimum due to
mixing, is a main subject for the following discussion.

Also shown in Fig.~\ref{alnidx} are estimates of $D_{cc}$ according to
the Darken equation, Eq.~\eqref{darken} with $S=1$ (open symbols). While
the concentration dependence is captured qualitatively correctly, the
Darken equation significantly overestimates $D_{cc}$. This effect is
more pronounced at higher temperatures, while the quantitative error
made by Eq.~\eqref{darken} is less at lower temperatures. To rationalize
this is a second main point for the discussion below.

\begin{figure}
\includegraphics[width=.9\linewidth]{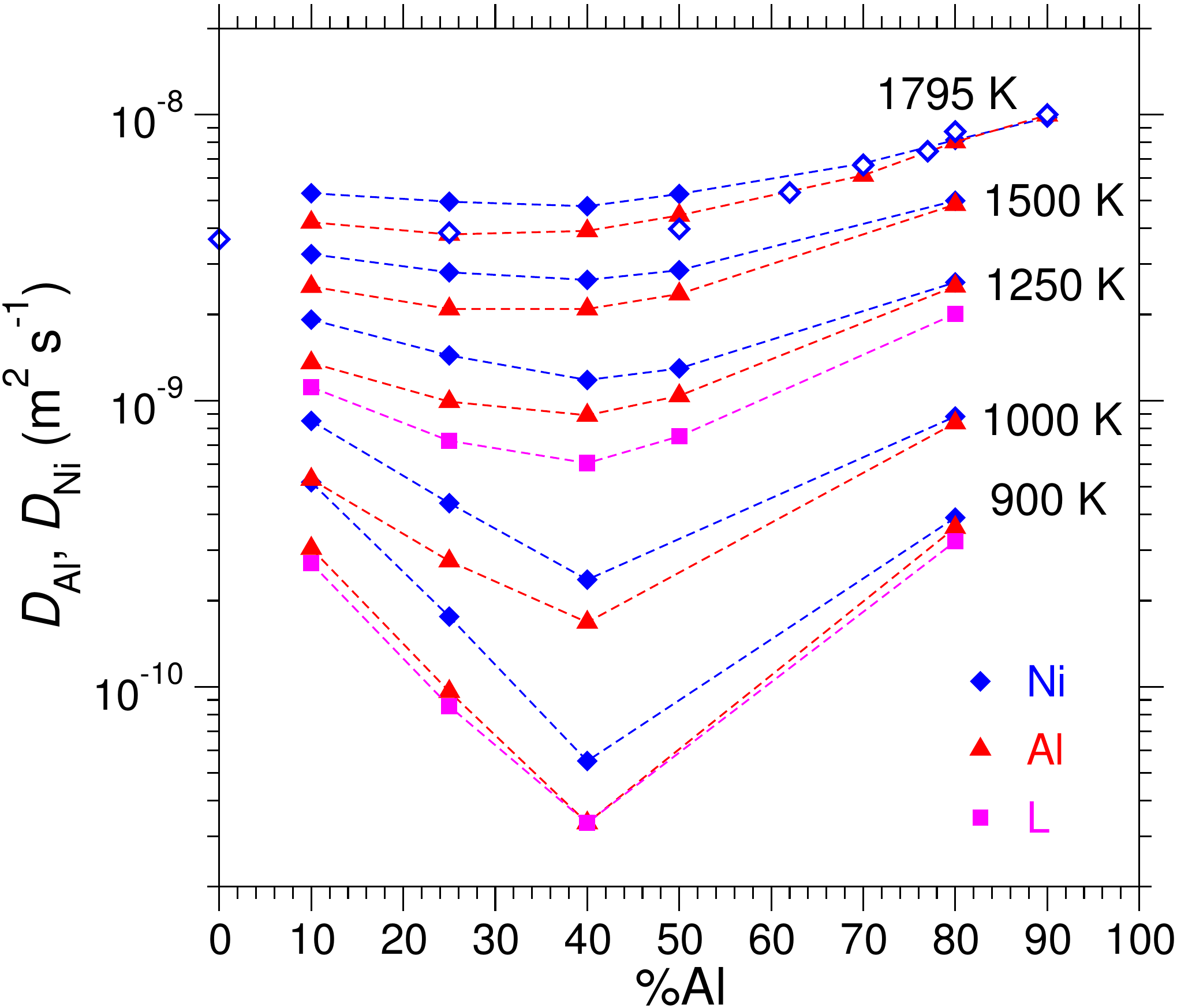}
\caption{\label{alnids}
  Al- (filled triangles) and Ni- (filled diamonds)
  self diffusion coefficients of the model Al-Ni system from
  MD simulations, as a function of Al concentration at various temperatures
  as labeled.
  For the temperatures $T=1250\,\text{K}$ and $900\,\text{K}$, also the
  Onsager coefficient $L$ for interdiffusion is shown (squares).
  Experimental data for $D^s_\text{Ni}$ at $T=1795\,\text{K}$ obtained
  from quasielastic neutron scattering (Ref.~\protect\onlinecite{Stueber.2010})
  are shown as open diamonds.
}
\end{figure}

Other than the interdiffusion coefficient,
the self-diffusion coefficients in the MD simulation always display a minimum
as a function of concentration at fixed temperature, in the whole range we
investigated.
This is shown in Fig.~\ref{alnids}.
As a generic trend, Ni diffusion is slightly faster, but on the
Al-rich side, the difference is much less pronounced.
For $T=1795\,\text{K}$, a shallow minimum in both self-diffusion coefficients
is seen on the Ni-rich side. Up to $x_\text{Al}\approx0.5$, the $D^s_\alpha$
depend only weakly on concentration: this is compatible
with recent experimental data on Ni-self
diffusion obtained from quasielastic neutron
scattering \cite{Stueber.2010} (open diamonds in
Fig.~\ref{alnids}). The MD simulation systematically overestimates
self-diffusion coefficients on the Ni-rich side. For example, for
$T=(1514\pm5)\,\text{K}$, a value of $D^s_\text{Ni}\approx(2.09\pm0.08)\,
\text{m}^2/\text{s}$ was measured in pure Ni using quasi-elastic
neutron scattering combined with electromagnetic levitation \cite{Meyer.2008};
the value estimated from Fig.~\ref{alnids} is about a factor of $3$ higher.
On the Al-rich side,
long-capillary experiments under microgravity conditions \cite{Garandet.2004}
obtained $D^s_\text{Ni}\approx3.7\times10^{-9}\,\text{m}^2/\text{s}$ in
(almost pure) $\text{Al}$ at $T=969\,\text{K}$.
This value is (if somewhat high) compatible with our MD results. Note however,
that uncertainties, e.g., relating to temperature control may be
considerable in this experiment.
The MD simulation extends the temperature window covered in the mentioned
experiments to considerably lower temperatures.
Upon decreasing the temperature, the intermediate-concentration minimum
in both self-diffusion coefficients becomes more pronounced and moves
towards higher Al-concentrations (about $x_\text{Al}\approx0.4$). At
$T=900\,\text{K}$, it amounts to about a factor $10$ decrease.

\begin{figure}
\includegraphics[width=.9\linewidth]{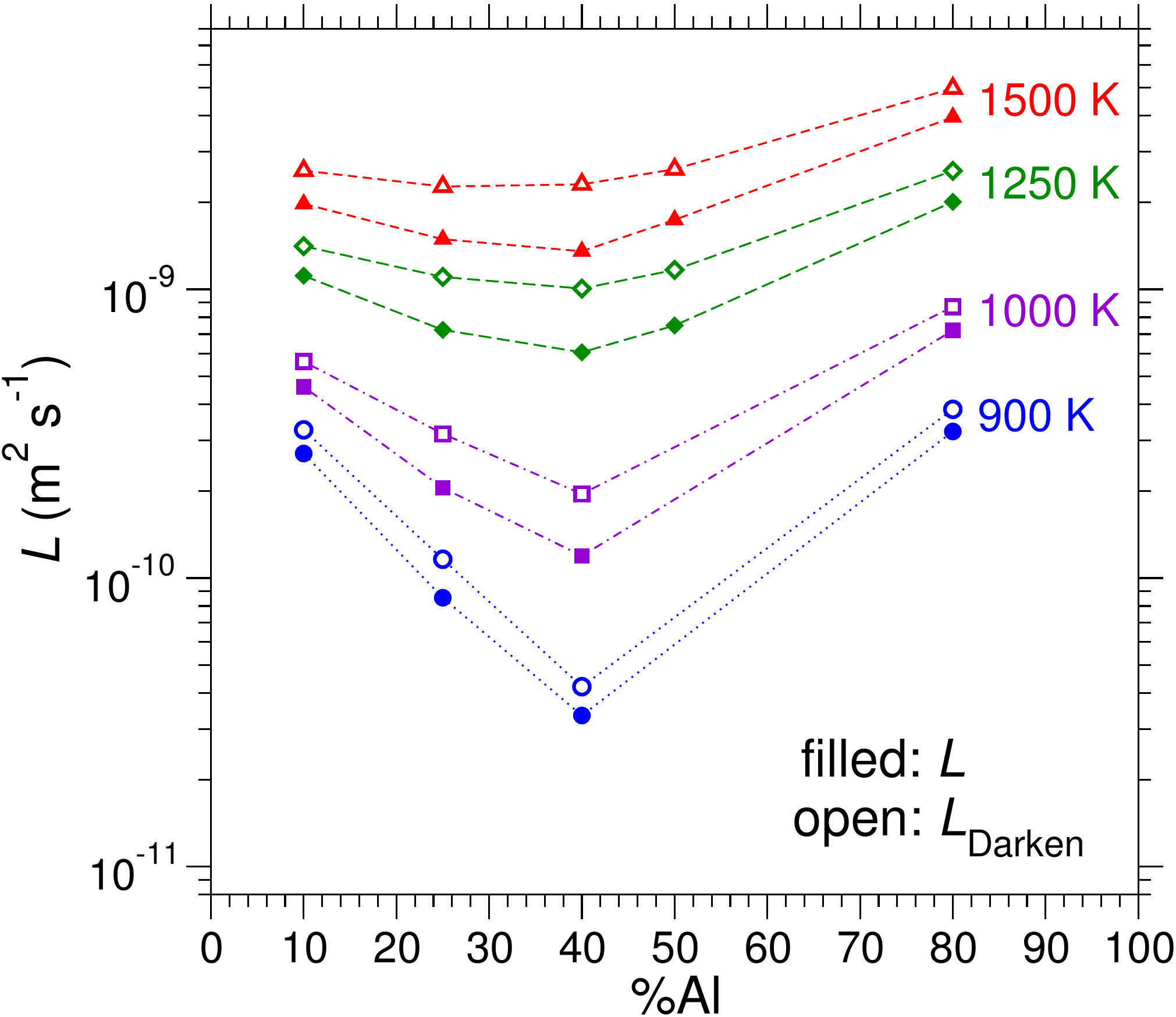}
\caption{\label{alnionsager}
  Onsager kinetic coefficient for interdiffusion, $L$, in the Al-Ni
  MD computer simulation, as a function of Al concentration at various
  fixed temperatures as indicated (filled symbols). Open symbols represent
  the estimate based on only the self-diffusion coefficients
  according to the Darken equation.
}
\end{figure}

The behavior of the self-diffusion coefficients already indicates the
mechanism responsible for the change of the concentration-dependence
of the interdiffusion coefficient from a maximum due to mixing, to a
minimum due to mixing. Recall that $D_{cc}$ is decomposed into a kinetic
part, the Onsager coefficient $L$, and a thermodynamic driving force,
$\Phi$. In the dense melt, one expects that qualitatively, all kinetic
factors behave similarly, because relaxation at high densities and/or low
temperatures is a strongly cooperative process.
Figure~\ref{alnionsager} shows the kinetic contribution to
$D_{cc}$ as filled symbols. At all temperatures, the Onsager coefficient
exhibits a minimum at intermediate concentrations. This minimum becomes
more pronounced upon lowering the temperature, similar to the trend already
observed for the self-diffusion coefficients. At the lowest temperature
investigated here, $T=900\,\text{K}$, the suppression of the interdiffusion
kinetics upon mixing is almost a factor of $8$, i.e., even more than the
suppression of the interdiffusion coefficient. From this observation,
one already infers that the thermodynamic factor $\Phi$ must display
a maximum as a function of concentration.
Again, the minimum in the kinetic coefficient $L$
is observed near the composition with $x\approx0.4$.

\begin{figure}
\includegraphics[width=.9\linewidth]{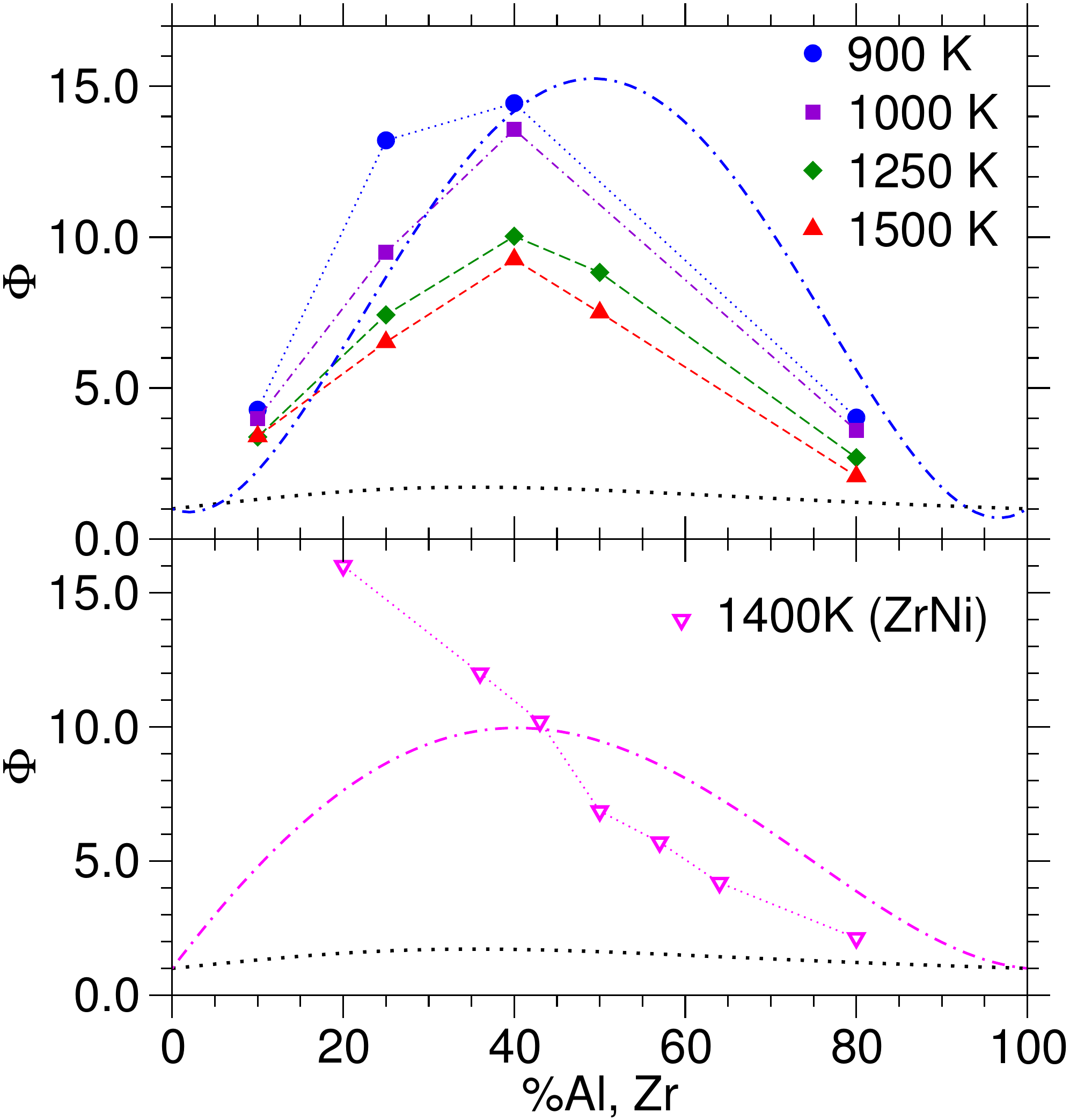}
\caption{\label{alnizrniphi}
  Thermodynamic factor $\Phi$ connecting the Onsager coefficient $L$ and
  the interdiffusion coefficient $D_{cc}$. Results from molecular-dynamics
  simulations for a model Al-Ni system are shown as filled symbols
  as functions of Al concentration (upper panel;
  constant temperatures as indicated).
  Open symbols (lower panel) are results from MD simulations of a model
  Zr-Ni system,
  as function of Zr concentration.
  Dotted lines indicate the pure entropic hard-sphere contribution
  expected for strongly different interaction radii, at constant density.
  Dash-dotted lines are the results of thermodynamic modeling
  (Refs.~\protect\onlinecite{Huang.1998} and \protect\onlinecite{Ghosh.1994})
  for Al-Ni at $T=900\,\text{K}$ respectively for Zr-Ni at $T=1400\,\text{K}$.
}
\end{figure}

The corresponding concentration dependence of the thermodynamic factor $\Phi$
is shown in Fig.~\ref{alnizrniphi}.
As expected from the above discussion, a maximum is found
for all temperatures, again at concentrations near $x=0.4$. For the lowest
temperature shown, an almost 15-fold increase with respect to the pure
systems is found.
The increase compared to the lowest $\text{Al}$-
respective $\text{Ni}$-concentrations investigated in our MD simulations,
is roughly a factor
of $3$. At higher temperatures, a similar maximum in $\Phi$ is seen, albeit
less pronounced.
The results from our MD simulation agree well with a recent \emph{ab initio}
molecular dynamics study for the $x=0.8$ composition \cite{Wang.2011}.

A standard empirical method to obtain thermodynamic factors is the so-called
CALPHAD method: based on large databases of experimentally determined
thermodynamic properties, suitable interpolation techniques are used to
reconstruct the Gibbs free energy $G$. For binary systems, interpolation
polynomials up to second order in $(x_\text{A}-x_\text{B})$ with linearly
$T$-dependent coefficients (so-called Redlich-Kister polynomials) are usually
employed.
For the Al-Ni liquid, parameters were determined by Huang and Chang
\cite{Huang.1998}. The resulting thermodynamic factor $\Phi$ is shown
in Fig.~\ref{alnizrniphi} as a dash-dotted line (for $T=900\,\text{K}$).
Agreement with our MD simulations is fair, and the magnitude of the
increase in $\Phi$ is captured correctly.
The interpolation assumed in
Ref.~\onlinecite{Huang.1998} implies that $\Phi(x)$ is almost symmetric around
$x\approx1/2$, and has its maximum there. Our MD simulation indicates
that the maximum in $\Phi(x)$ is shifted somewhat to the Ni-rich side,
as was already noted for a higher temperature earlier \cite{Griesche.2009}.
Furthermore for Al-Ni, the CALPHAD calculations suggest $\Phi<1$ if one of
the components is very dilute. The investigation of this parameter regime
in experiment and simulation is the subject of a separate publication
\cite{Sondermann_pre}.
Note also that the parametrization of Ref.~\onlinecite{Huang.1998} is
constructed to work in the stable liquid phase, while we use it to
smoothly extrapolate to the metastable supercooled liquid.

The overall temperature variation of the thermodynamic factor at a fixed
concentration is weaker than that of the Onsager coefficient. In addition,
it is reversed: while kinetic coefficients like $L$ or the self-diffusion
coefficients drop sharply with decreasing temperature, the thermodynamic
factor increases.
Taken together, at high temperatures the maximum in the thermodynamic
factor dominates the behavior of $D_{cc}=L\cdot\Phi$, so that a maximum
in the interdiffusion coefficient is observed. At low temperatures,
the stronger suppression and more pronounced minimum in $L$ dominates,
resulting in a minimum in the interdiffusion coefficient although the
maximum in the thermodynamic factor becomes more pronounced.

That the thermodynamic factor exhibits a maximum is expected to be a rather
generic effect: deviations from ideal-gas-like mixing in a thermodynamically
stable system that favors mixing
will generate thermodynamic driving forces acting to
level out concentration fluctuations. This corresponds to a thermodynamic
factor $\Phi>1$.
Such a trend is even observed without any attractive interactions and
on purely entropic grounds: already the hard-sphere system displays a
mixing-induced maximum in $\Phi$. Although more accurate empirical
equations of state are available for hard-sphere mixtures, the effect
is indicated in the Percus-Yevick (PY) approximation for the static structure
factor \cite{Hansen}. Since PY approximates the system to be stable
and mixing for all parameters, one can discuss its thermodynamic factor
for simplicity in the limiting case of vanishing size of
one of the species.
Evaluating $\Phi=x(1-x)/S_{cc}(q\!=\!0)$ with $x$ indicating the
large-particle number concentration, one gets for size ratio $\delta\ll1$
\begin{equation}
  \Phi_\text{HSPY,$\delta\to0$}=\frac{\left(1+\frac{\pi}{3}x\rho\right)^2}
  {1+\frac{\pi}{6}x\rho(6x+\frac{\pi}{6}x\rho(1+3x)-2)}\,.
\end{equation}
This result is shown as a dotted line in Fig.~\ref{alnizrniphi}. It
exhibits a maximum value of approximately $1.7$ at a concentration
$x\approx0.35$. The observed thermodynamic factors in metallic
melts are typically much larger, which may be intuitively explained by the
lack of inter-species attractions and non-additive mixing leading to a much
suppressed chemical ordering in hard-sphere mixtures \cite{Voigtmann.2008b}.
It is nevertheless remarkable that the maximum position for $\Phi$ at
concentrations in the range around $40\%$ of the larger particles may be
rather generic whenever the system is dense enough so that the
core-repulsion contribution from the interparticle interactions becomes
significant.

As will be discussed below in connection with MCT, a strong decrease
in the kinetic coefficients due to mixing is also expected to be rather
generic in dense systems, as it also arises already in the hard-sphere
mixture model. The evolution of the concentration dependence in the
interdiffusion coefficient highlighted in Fig.~\ref{alnidx} should
therefore be typical for dense metallic melts, given that the relevant
temperature and concentration ranges can be explored without intervening
phase separation, crystallization, or other effects that prevent the
(supercooled) liquid state to be accessed.

\begin{figure}
\includegraphics[width=.9\linewidth]{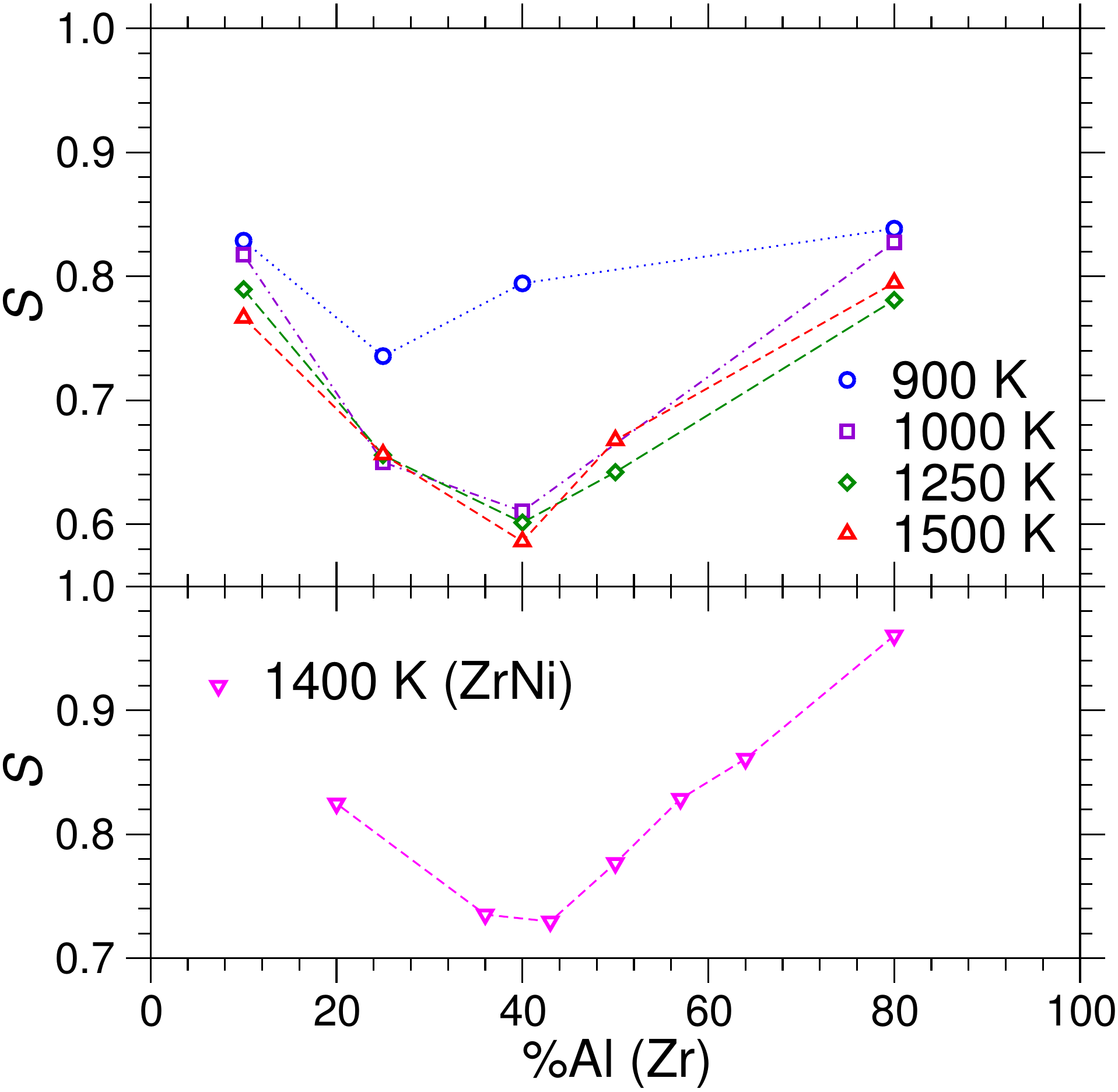}
\caption{\label{alnizrnimanning}
  Correction factor $S$ for the Darken equation obtained from MD computer
  simulation of Al-Ni and Zr-Ni models, as a function of Al respective Zr
  concentration, for the temperatures indicated.
}
\end{figure}

By equating the Onsager coefficient of interdiffusion to the weighted
sum of the self-diffusion contributions, the Darken equation expresses the
assumption that all kinetic contributions to diffusion processes share the
same physical mechanism. In the MD simulation, Darken's assumption is easily
tested. Figure~\ref{alnionsager} includes as open symbols the Darken
average. It is seen that the Darken equation in Al-Ni systematically
overestimates the Onsager coefficient. This is also evident already
from Fig.~\ref{alnids}, where the Onsager coefficient $L$ is shown for
two exemplary temperatures. It is smaller than either of the self-diffusion
coefficients, in contradiction to Darken's Eq.~\eqref{darken} with $S=1$.
The qualitative trend of
the Onsager coefficient both with changing temperature and changing
concentration is captured by the Darken approximation. This is
a consequence of the fact that all kinetic, non-thermodynamic, contributions
to the diffusion coefficients slow down with decreasing temperature and
with concentration approaching $x\approx40\%$.

The quality of the Darken approximation is best visualized in terms of
the correction factor $S=L/(x_\text{B}D^s_\text{A}+x_\text{A}D^s_\text{B})$.
This is shown in Fig.~\ref{alnizrnimanning}. As already evident from
Fig.~\ref{alnionsager}, $S\le1$ holds for all compositions and temperatures
investigated in our Al-Ni simulations. At high temperatures and intermediate
concentrations, $S\approx0.6$, i.e., the Darken approximation overestimates
the interdiffusion coefficient by about $66\%$. By construction, the
agreement becomes better if either $x\approx0$ or $x\approx1$ is
approached. There, $S=1$ should hold on theoretical grounds.
At the lowest temperature, $T=900\,\text{K}$, the correction factor
is significantly closer to unity at all concentrations, obeying $S\approx0.8$
(corresponding to a $25\%$ over-estimation by Darken's approximation).
This is in line with the expectation that the lower the temperature, the
stronger the cooperative relaxation effects that imply a tight coupling
between self- and collective diffusion processes. As explained below,
this is the picture emerging from mode-coupling theory as the temperature
is lowered towards the kinetic glass transition.
According to our MD simulatons, there is a marked increase in $S$ between
$T=1000\,\text{K}$ and $T=900\,\text{K}$. Note that coincidentially,
the concentration dependence of the $D_{cc}$ starts to be dominated by
kinetic effects around $T=900\,\text{K}$ (it displays a minimum as a function
of concentration as seen in Fig.~\ref{alnidx}), while it is still dominated
by thermodynamic effects at $T=1000\,\text{K}$.

\subsection{Zr-Ni Simulations}

\begin{figure}
\includegraphics[width=.9\linewidth]{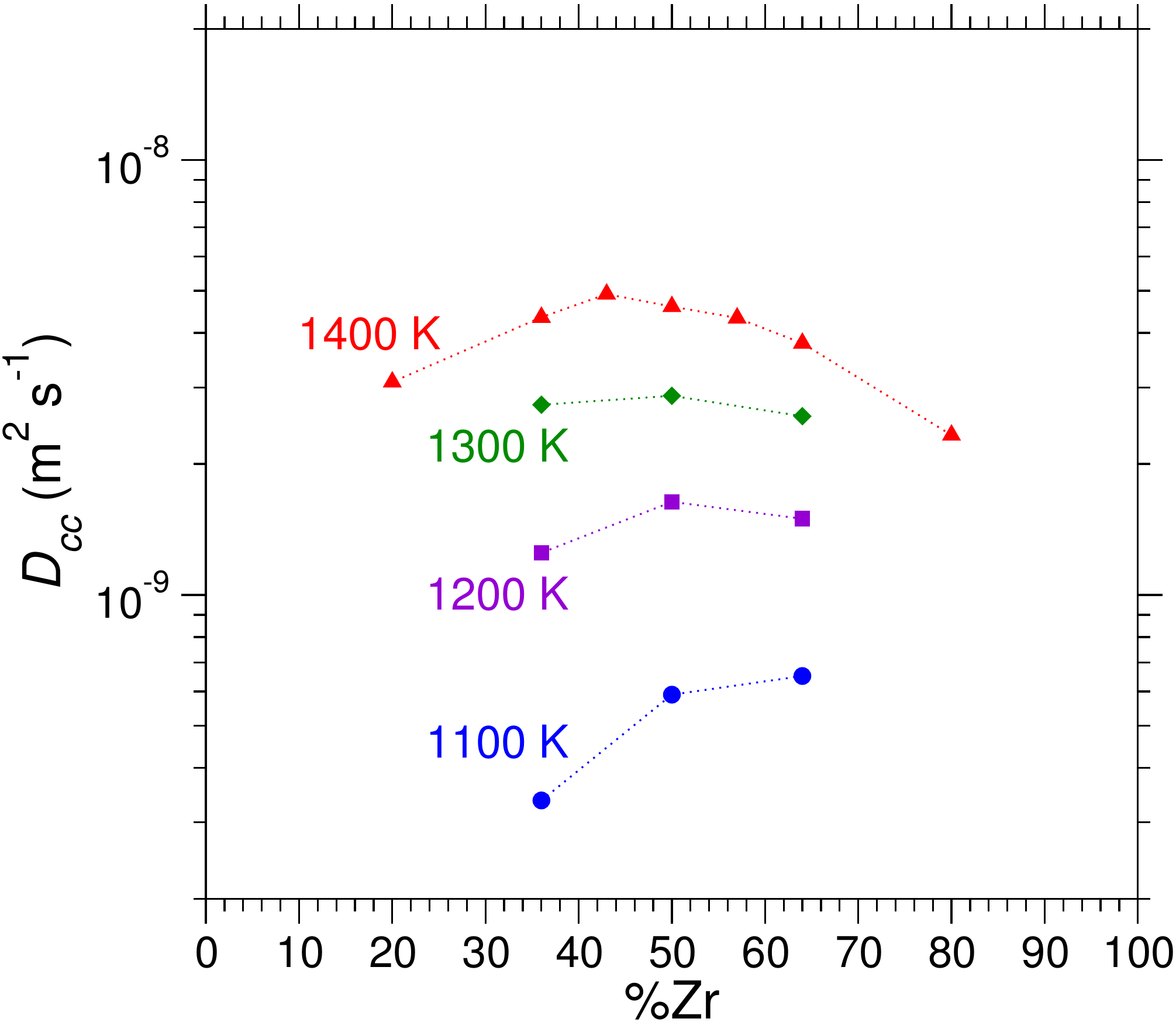}
\caption{\label{zrnidx}
  Interdiffusion coefficients $D_{cc}$ of a Zr-Ni model system from
  MD computer simulation, at various fixed temperatures as indicated and
  as a function of Zr-concentration.
}
\end{figure}

To put our findings for the Al-Ni model in context, we next discuss
results from molecular-dynamics simulations on a Zr-Ni model system.
As mentioned above, these two binary mixtures display quite different
thermodynamic phase diagrams, and are representatives for mixtures with
different effective-size ratios in terms of their kinetics.
However, the qualitative arguments given above in the case of Al-Ni,
are expected to be generic, so that a test for Zr-Ni as a second
representative example is in order.

Figure~\ref{zrnidx} shows results for the interdiffusion coefficient
$D_{cc}$ to be compared to the corresponding Al-Ni result in Fig.~\ref{alnidx}.
Keeping in mind that for the Zr-Ni model, MD simulations were only carried
out for $x\ge0.36$ at low temperatures, one indeed observes qualitative
similarities between the two systems. For high temperatures, $D_{cc}$
in the Zr-Ni model decreases displays a maximum around $x\approx0.4$.
At the lower temperatures, the MD data do not allow to draw a definite
conclusion, but they are compatible with the reduction of this maximum
and the development of a minimum at some $x\lesssim0.4$.

\begin{figure}
\includegraphics[width=.9\linewidth]{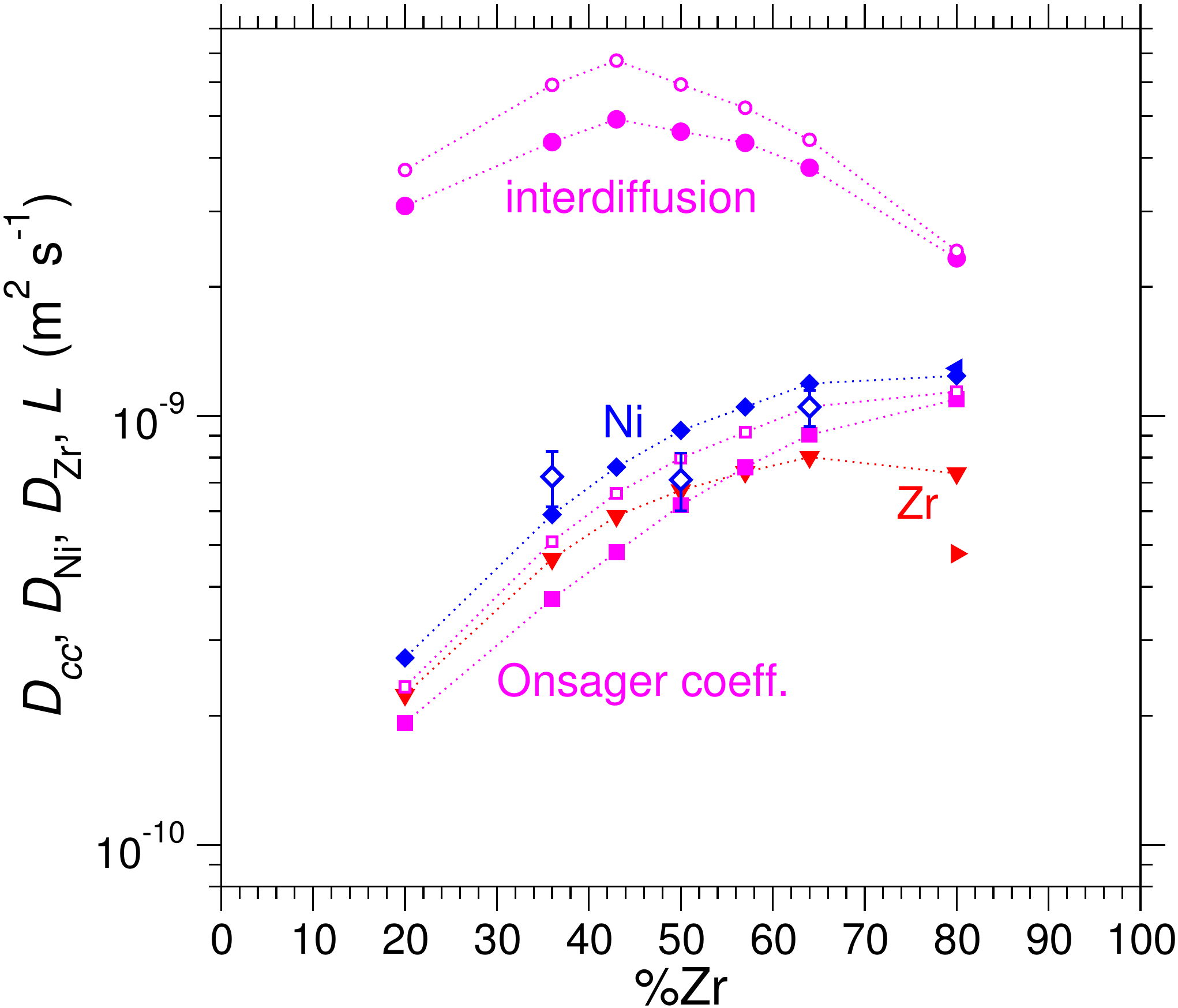}
\caption{\label{zrnixonet}
  Interdiffusion coefficient $D_{cc}$ (filled circles), related Onsager kinetic
  coefficient $L$ (filled squares), and self-diffusion coefficients for
  Ni and Zr (filled diamonds and inverted triangles),
  for the Zr-Ni simulation model at fixed temperature
  $T=1400\,\text{K}$, as a function of Zr concentration.
  Open squares and circles are the Onsager coefficient and $D_{cc}$
  estimated from the Darken equation.
  Left- and right-pointing triangles are the self-diffusion coefficients
  for Ni and Zr, obtained using a
  different MD simulation model from Ref.~\protect\onlinecite{Mutiara.2001}.
  Open diamonds with error bars
  are experimental data for $D_\text{Ni}$ from quasielastic
  neutron scattering (interpolated from
  Ref.~\protect\onlinecite{HollandMoritz.2009}).
}
\end{figure}

In Fig.~\ref{zrnixonet}, we show all diffusion coefficients, $D_{cc}$,
$D^s_\text{Ni}$, and $D^s_\text{Zr}$, evaluated from the Zr-Ni MD simulation
model at the fixed temperature $T=1400\,\text{K}$. In difference to the
Al-Ni simulation discussed above, the self-diffusion coefficients show a
notable increase with increasing Zr concentration.
Experimental values for $D^s_\text{Ni}$ have been obtained by
quasielastic neutron scattering and compiled in
Ref.~\onlinecite{HollandMoritz.2009}. Agreement with the MD simulation
is reasonable, but not surprising since the simulation potential has been
tuned based on the diffusion data in $\text{Zr}_\text{64}
\text{Ni}_\text{36}$. Experimental values for $\text{Zr}_\text{50}
\text{Ni}_\text{50}$ are noticeably lower than expected from the simulation;
the reason for this discrepancy is unclear.
For $x_\text{Zr}=0.36$, the experimental value is above the one calculated
from the simulation. Recall that for pure Ni, experiments
\cite{Meyer.2008} suggest a value $D^s_\text{Ni}\gtrsim1\times10^{-9}\,
\text{m}^2/\text{s}$, significantly higher than the MD-simulated value
for the lowest Zr-concentration that is accessible without crystallization.
This may suggest that the Zr-Ni model employed in our MD simulations
underestimates the diffusivities on the Ni-rich side. On the other hand,
the model suggests strong concentration dependence of some quantities in
the range $x_\text{Zr}\le0.2$, as will be discussed below in connection with
the thermodynamic factor shown in Fig.~\ref{alnizrniphi}.
In this case, a pronounced minimum in $D^s_\alpha$ would occur in this
concentration regime.

At all concentrations,
one has $D^s_\text{Zr}<D^s_\text{Ni}$, which is expected from the simple
analogy to hard-sphere mixtures since $\text{Ni}$ is the ``smaller'' atom.
Previous calculations using MCT based on static structure factors
determined by neutron scattering \cite{Voigtmann.2008b} suggested that
$D^s_\text{Zr}\approx D^s_\text{Ni}$ holds at least for $T=1350\,\text{K}$
and $x_\text{Zr}=0.64$. These static structure factors showed a strong
pre-peak indicating strong chemical short-range order that suppresses the
decoupling of Ni and Zr diffusion one expects from the purely entropic
hard-sphere mixture. In our MD simulation model, this pre-peak is much
less pronounced. The enhancement of $D^s_\text{Ni}$ over $D^s_\text{Zr}$
in the simulation increases with increasing Zr concentration, up to
$D^s_\text{Ni}\approx1.7D^s_\text{Zr}$ at $x_\text{Zr}=0.8$. This
compares reasonably well with the factor $2.5$ reported for a different
Zr-Ni model by Mutiara and Teichler \cite{Mutiara.2001}. Values from
this study are shown in Fig.~\ref{zrnixonet} as left- and right-pointing
triangle symbols.

Again, for the Zr-Ni system the MD simulation allows to separate the
thermodynamic factor $\Phi$ and the correction factor $S$ from
the overall interdiffusion coefficient. Concentration dependent results
at fixed temperature from our Zr-Ni simulations are shown together with
the Al-Ni results in Figs.~\ref{alnizrniphi} and \ref{alnizrnimanning}.
One confirms that both terms are of equal magnitude in the two systems.
Again, the correction factor $S$ displays a minimum for concentrations
corresponding to roughly $60\%$~Ni. At the same time, the values for
$S$ obtained in the Zr-Ni simulation are somewhat higher than in Al-Ni
at the same temperature,
approaching unity on the Zr-rich side for $T=1400\,\text{K}$.
Here one has to keep in mind that at equal temperatures, the Zr-Ni
system is kinetically slower, as evidenced from Figs.~\ref{alnids} and
\ref{zrnixonet}. Assuming that slow cooperative
relaxation processes are responsible for a strong coupling between
all modes of diffusion, and hence drive $S$ towards unity, this rationalizes
the better quantitative (and similar qualitative) performance of Darken's
approximation in the Zr-Ni melt.

The thermodynamic factor in our Zr-Ni model at $T=1400\,\text{K}$
monotonically increases over the concentration range covered
in the simulations ($x\ge0.2$). Since $\Phi\to1$ should hold for $x\to0$,
we conclude that the thermodynamic factor displays a pronounced maximum
in the Zr-concentration interval $x_\text{Zr}\in[0,0.2]$. Rapid crystallization
in the simulation prevents us from checking this directly.
Such a maximum is at significantly lower concentration than the maximum
in $\Phi$ found in the Al-Ni system. The estimate based on interpolation
of thermodynamic property data, with parameters for Redlich-Kister polynomials
obtained by Ghosh \cite{Ghosh.1994} (shown in Fig.~\ref{alnizrniphi} as
a dash-dotted line), predicts the maximum to be shifted to the Ni-rich
side compared to Al-Ni, albeit not as much as suggested by the MD simulation.
The reason for the strong increase observed in the simulation is not clear.
We could not find obvious structural changes in the melt compared to
other concentrations.

\begin{figure}
\includegraphics[width=.9\linewidth]{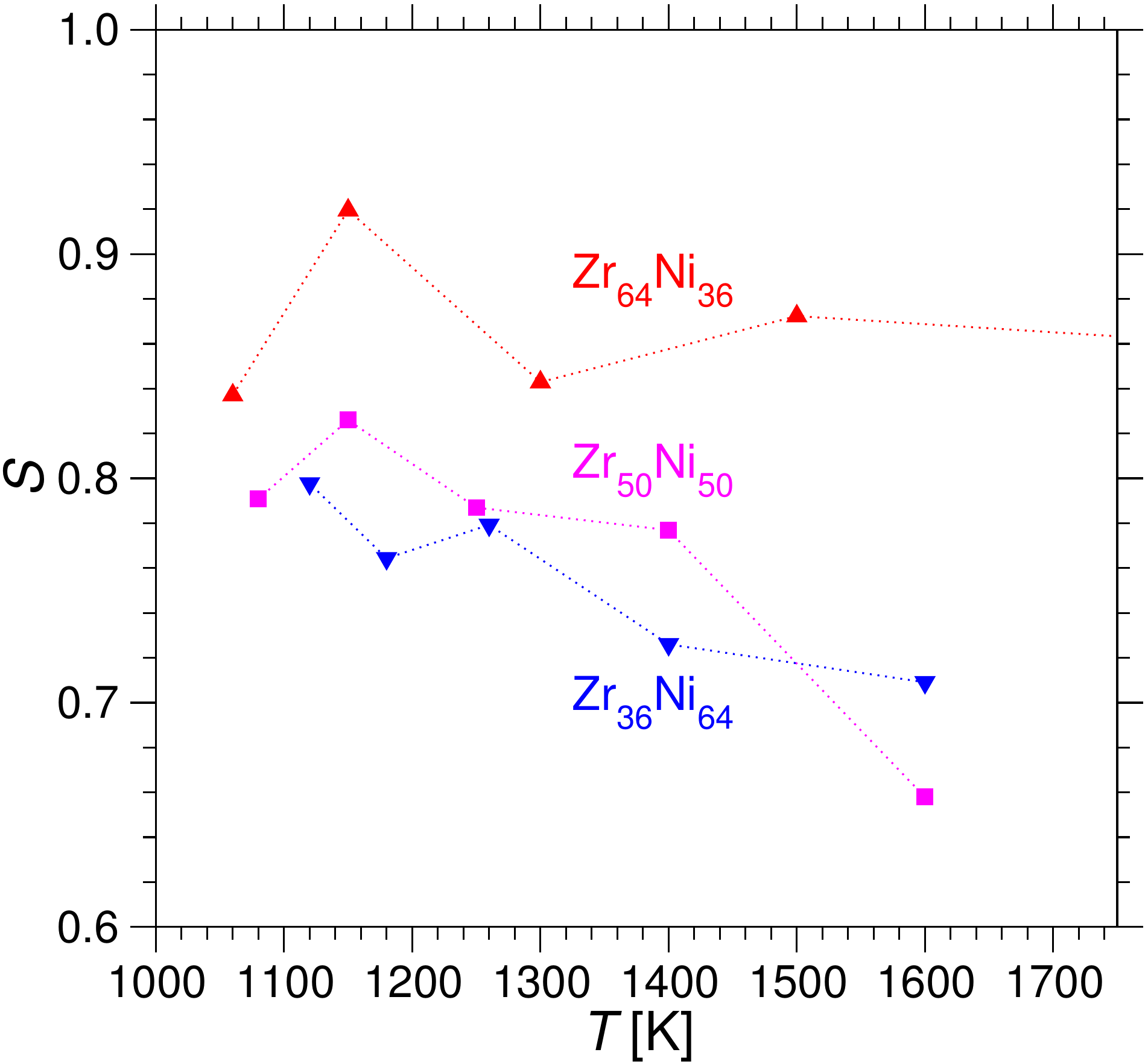}
\caption{\label{zrnis}
  Correction factor $S$ for the Darken equation for the interdiffusion
  coefficient in the Zr-Ni MD simulation model for three different
  compositions, as a function of temperature.
}
\end{figure}

\begin{figure}
\includegraphics[width=.9\linewidth]{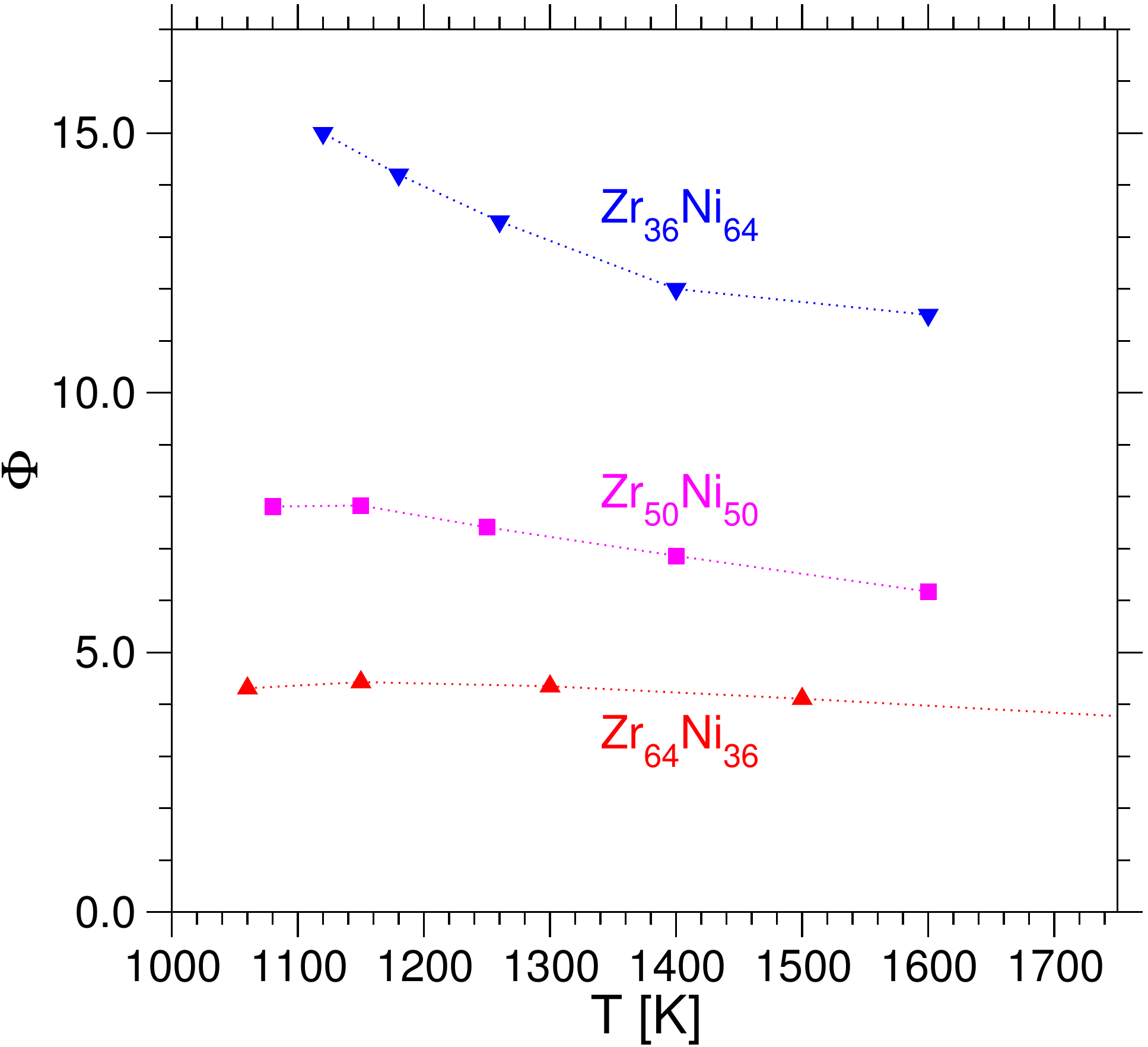}
\caption{\label{zrniphi}
  Thermodynamic factor $\Phi$ for interdiffusion in the Zr-Ni
  model for three different compositions as labeled, as a function
  of temperature.
}
\end{figure}

For the Zr-Ni system, we also show the temperature dependence of the
correction factor $S$ and the thermodynamic factor $\Phi$, in Fig.~\ref{zrnis}
respectively \ref{zrniphi}, at the three compositions $x=0.36$, $x=0.5$,
and $x=0.64$ for which a large temperature window was studied in our
simulations. Turning first to the dynamic correction factor $S$,
we find a week increase with decreasing temperature indicating the
crossover from typical liquid-state dynamics at high temperatures to
a cage-effect dominated regime (where one expects $S$ to approach unity).
However, it has to be noted that the temperature variation of $S$ is rather
weak, and in particular for the $\text{Zr}_\text{64}\text{Ni}_\text{36}$
system, no obvious trend can be seen within the error bars of our simulation.
This is similar to what has been observed in $\text{Al}_\text{80}\text{Ni}_
\text{20}$ over a similar temperature range \cite{Horbach.2007}.

The thermodynamic factor, Fig.~\ref{zrniphi}, increases with decreasing
temperature as naively expected. This increase is more pronounced for
the $\text{Zr}_\text{36}\text{Ni}_\text{64}$ system, i.e., closer to the
composition where a maximum in $\Phi$ is expected as a function of
composition. Overall, the values of $\Phi$ extracted from our simulation
are compatible with those estimated from experiment \cite{Voigtmann.2008b}.
They are somewhat lower than the large thermodynamic factor reported
for the amorphous $\text{Zr}_\text{43}\text{Ni}_\text{57}$ solid
\cite{Karpe.1995}, which is however to be expected since we are here
dealing with the liquid state at much larger temperatures.

\subsection{MCT Results}

The development of a pronounced minimum in the kinetic transport coefficients
at intermediate concentrations and low enough temperatures can be rationalized
within the mode-coupling theory of the glass transition.
It has already been noted that for hard-sphere mixtures of not too disparate
size, MCT predicts a favorization of the glass by mixing \cite{Goetze.2003}.
The total packing fraction at the kinetic glass transition in this regime
is decreased by mixing, such that along a cut of constant packing fraction,
increasing the minority-species concentration towards roughly $x\approx1/2$
will decrease the control-parameter distance to the glass transition.
This translates into a suppression of kinetic transport coefficients
as a function of species concentration at fixed packing fraction, as
confirmed in event-driven MD simulations of binary hard-sphere mixtures
\cite{Foffi.2003}.
This suppression holds for size ratios $0.75\lesssim\delta<1$ and arises
from the increased structural disorder introduced by the minority species
into the pure system.
It is remarkable since for colloidal suspensions, where mixtures with
stronger size disparity are typical, the opposite effect is usually
discussed and rationalized by effective interactions induced entropically
between the large-species particles \cite{Voigtmann.2011b}.
In this case, the kinetic coefficients typically show a maximum at intermediate
concentrations and fixed volume fraction, equivalent to a strong fluidization
of the system by mixing. However, at size ratios close to unity, the
depletion-interaction effect is no longer the dominant mechanism, since here
the minority species makes a strong contribution to the static structure
that is not accounted for in the depletion picture.

\begin{figure}
\includegraphics[width=.9\linewidth]{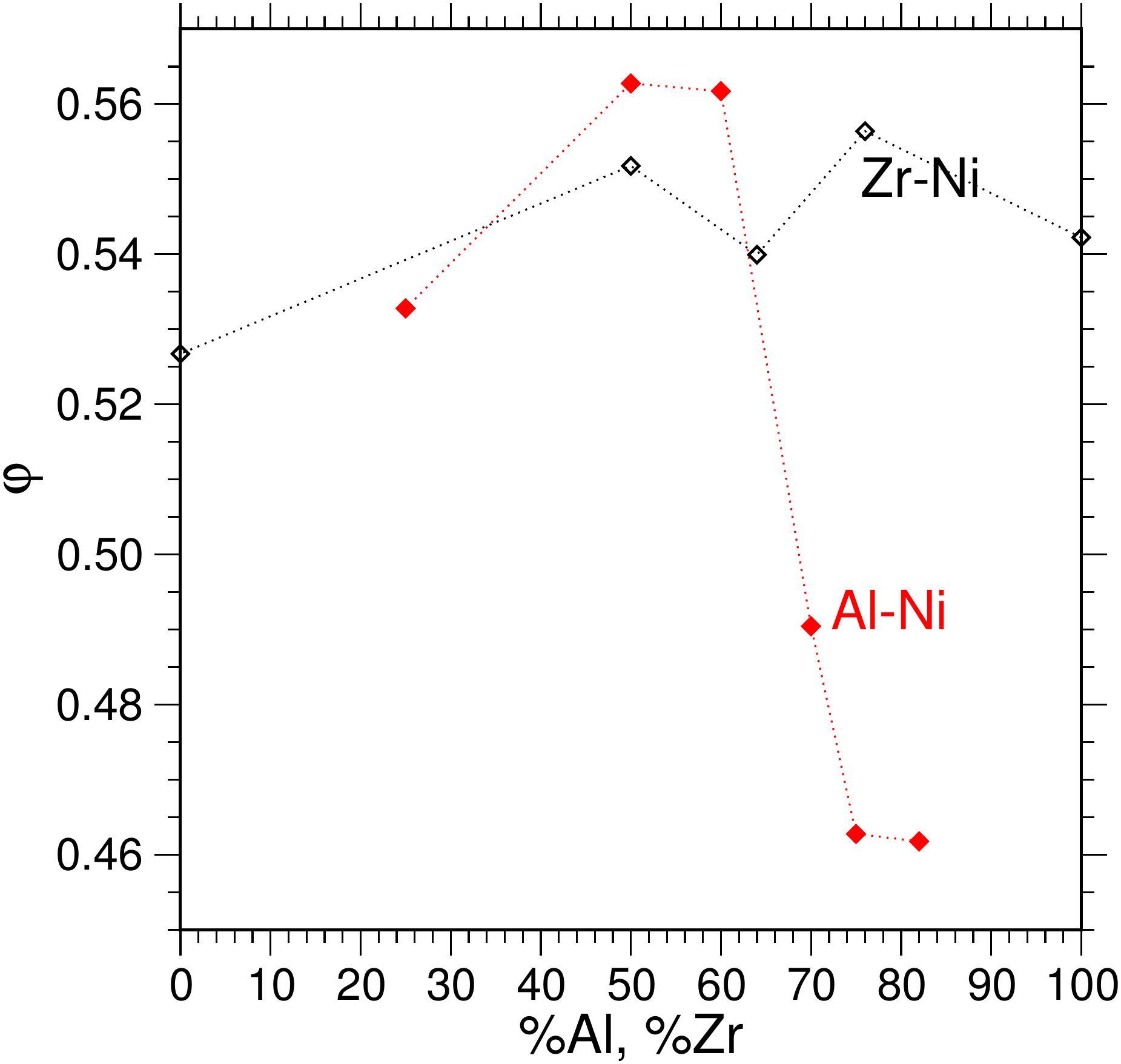}
\caption{\label{pfr}
  Effective packing fractions $\varphi$ estimated for various Al-Ni (filled
  symbols) and Zr-Ni (open symbols) melts, at $T=1473\,\text{K}$ for various
  compositions.
}
\end{figure}

A rough estimate of entropic size effects in metallic melts is provided
by the covalent radii of the species. Estimated values \cite{Pauling.1947} are
$R_\text{Zr}=1.45\,\text{\AA}$, $R_\text{Ni}=1.15\,\text{\AA}$, and
$R_\text{Al}=1.25\,\text{\AA}$. From this, size ratios are evaluated
(including error bars obtained from different estimates of the radii
based on crystallography \cite{Cordero.2008},
$R_\text{Zr}=1.75\,\text{\AA}$, $R_\text{Ni}=1.24\,\text{\AA}$, and
$R_\text{Al}=1.21\,\text{\AA}$):
$\delta_{\text{Al}-\text{Ni}}\approx0.92\pm0.06$, and
$\delta_{\text{Zr}-\text{Ni}}\approx0.79\pm0.08$.
Based on this estimate, the kinetic slowing down upon mixing is indeed
expected in both systems upon entropic grounds, since $\delta\gtrsim0.75$
(as is the case for metallic melts quite generally).

Assuming the constituents of the dense melt to be reasonably close to
hard spheres, one can assign to it an effective packing fraction.
\cite{Meyer.2002}. Based on the above diameters and
the relative atomic
masses, $m_\text{Zr}=91.224\,\text{amu}$,
$m_\text{Ni}=58.693\,\text{amu}$, and $m_\text{Al}=26.982\,\text{amu}$,
the packing fraction $\varphi$ can be estimated from existing
experimental data for the mass density. The results for Al-Ni and
Zr-Ni are shown in Fig.~\ref{pfr}. Here, mass-density data have been
taken from the literature \cite{Plevachuk.2007,Paradis.2014} and
(in the case of Zr-Ni) extrapolated to constant temperature.
For the case of Al-Ni, one notices a marked maximum around
$x_\text{Al}\approx0.5$. This agrees qualitatively with the above result,
that the dynamics is slowest around this composition: for the
estimated size ratio, the hard-sphere packing fraction corresponding to
the MCT glass transition slightly decreases, while at the same time
the effective packing fraction at constant temperature increases. Both these
effects suggest that, entropically, the system is closer to arrest for
$x\approx0.5$ than for other compositions. For Zr-Ni, the
data do not allow to identify a clear trend.
However, the effective packing fractions in Zr-Ni are larger than those
of Al-Ni especially on the Ni-poor side; this agrees with our finding
that the dynamics in Zr-Ni is slower than that in Al-Ni at the same
temperature.

\begin{figure}
\includegraphics[width=.9\linewidth]{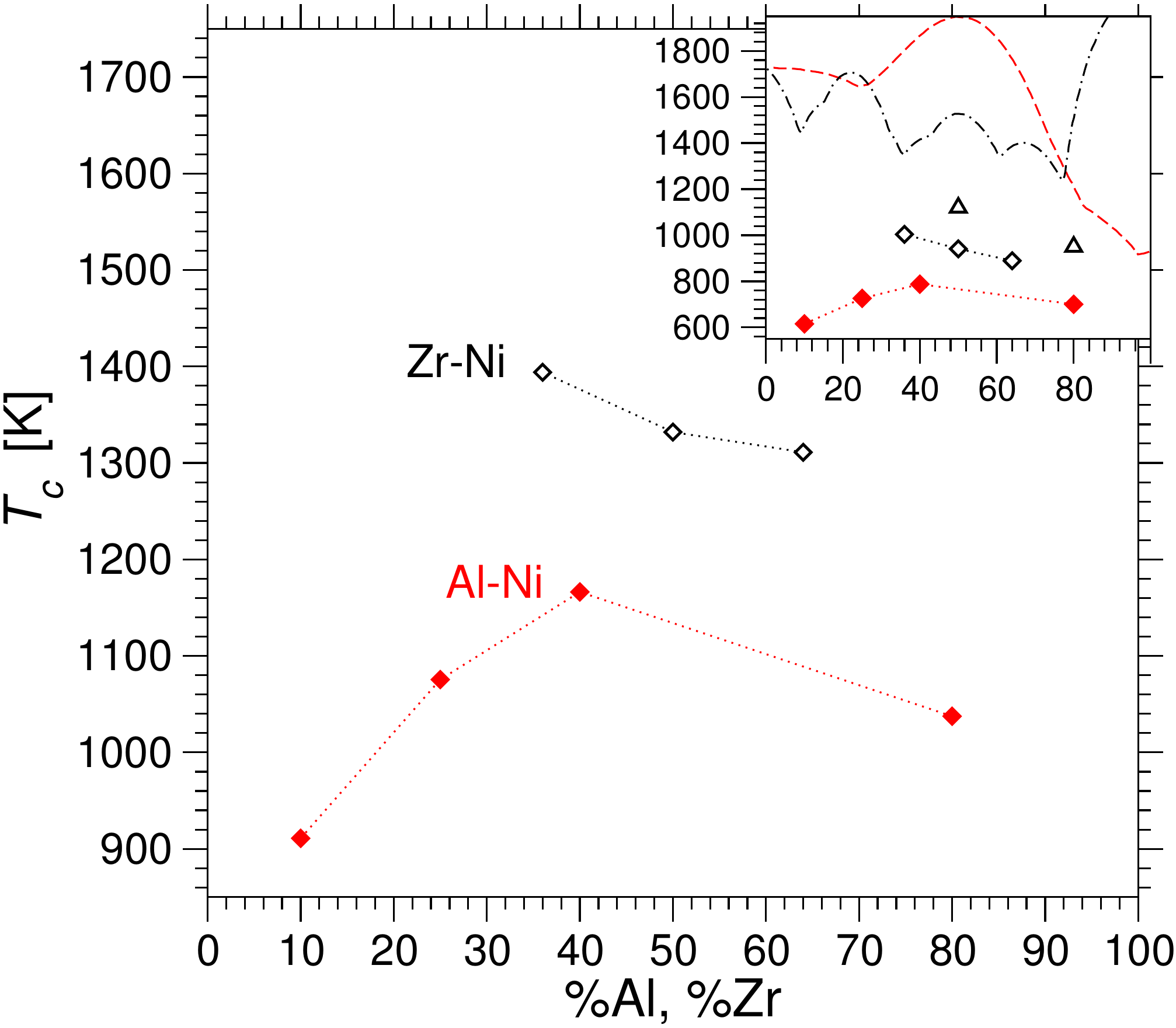}
\caption{\label{alnimcttc}
  Kinetic glass transition temperature $T_c$ for Al-Ni (filled symbolds)
  and Zr-Ni (open symbols)
  as a function of composition, obtained from mode-coupling
  theory of the glass transition (MCT) using the static
  structure factors from MD simulation.
  The inset shows the estimates of $T_c^\text{MD}$ from the MD simulation
  directly (see text for details), together with boundaries for the
  thermodynamic-equilibrium liquid phase (Al-Ni: dashed; Zr-Ni: dash-dotted).
  Triangles mark the $T_c^\text{MD}$ values obtained from a different
  Zr-Ni potential in Refs.~\protect\onlinecite{Teichler.1996,Mutiara.2001}.
}
\end{figure}

Without resorting to the hard-sphere analogy, one can for the MD-simulated
metallic melts directly test the
change of the MCT glass-transition temperature $T_c$ as a function
of composition. This is shown in Fig.~\ref{alnimcttc} for the Al-Ni model
system. $T_c(x)$ exhibits a maximum at intermediate $x$, so that along an
isotherm above $T_c$, the distance to the glass transition is decreased
upon approaching these intermediate-$x$ compositions.
This is in analogy to the discussion of the hard-sphere mixtures
above, and thus appears
to be driven by mainly entropic effects.

We also show in Fig.~\ref{alnimcttc} the values obtained from a MCT
calculation using the Zr-Ni static structure factor from our MD simulation.
In line with the finding that the dynamics of Zr-Ni is slower than that
of Al-Ni at the same temperature, the $T_c$ values for Zr-Ni are systematically
higher. Since $x=0$ in both simulations corresponds to a pure Ni system,
the corresponding $T_c(0)$ values should agree. In light of this, the
$T_c(x)$ values for Zr-Ni will also display a maximum for some
$x\le0.36$.

The $T_c$ values predicted by MCT are not accurate, and typically too high
since the theory overestimates the tendency to vitrify. This is a known
issue in comparing MCT calculations with MD simulation data
\cite{Sciortino.2001}. From the MD simulation, $T_c$ can be independently
estimated by analyzing the intermediate scattering functions in terms of
the asymptotic scaling laws provided by MCT \cite{Goetze.2009}. For
the Zr-Ni model employed here, the resulting estimates $T_c^\text{MD}$
are roughly a factor $1.4$ lower than the calculated $T_c^\text{MCT}$.
A similar difference has been observed earlier for the Al-Ni model
\cite{Das.2008}. The $T_c^\text{MD}$ for the two systems are shown in
the inset of Fig.~\ref{alnimcttc}. There, values obtained using a different
Zr-Ni model by Teichler and coworkers \cite{Teichler.1996,Mutiara.2001} are
added. These values are compatible with the same mixing trend, but are
systematically higher. The latter effect is due to our adaption of the
EAM potential to match experimentally measured diffusivities.
Let us stress that the MCT glass-transition
temperature $T_c$, being purely kinetic in origin, does not mirror
any equilibrium phase transition. In the inset of Fig.~\ref{alnimcttc},
dashed (dash-dotted) lines indicate the boundaries of the thermodynamically
stable liquid phase for Al-Ni (Zr-Ni), obtained from a standard
database \cite{mtbase}. One readily observes that the concentration dependence
of $T_c$ is not correlated with that of thermodynamic solid--liquid boundary.

Let us add a further remark concerning the idealized MCT glass transition.
According to MCT,
the diffusion coefficients approach zero as $T$ approaches $T_c$ from above,
and the viscosity diverges. This is the case for the ideal glass, which is
not attained in real-world experiment, due to residual relaxation processes
that are not captured in the MCT approximation. Thus, transport coefficients
remain finite even at $T_c$ and below. Nevertheless, a strong non-Arrhenius
decrease of kinetic coefficients is typically observed in glass formers
as one approaches $T_c$ from high temperatures. In this sense, the
MCT predictions discussed here are not quantitatively accurate, but
serve to explain the qualitative features of the slow dynamics in dense
metallic melts.

\begin{figure}
\includegraphics[width=.9\linewidth]{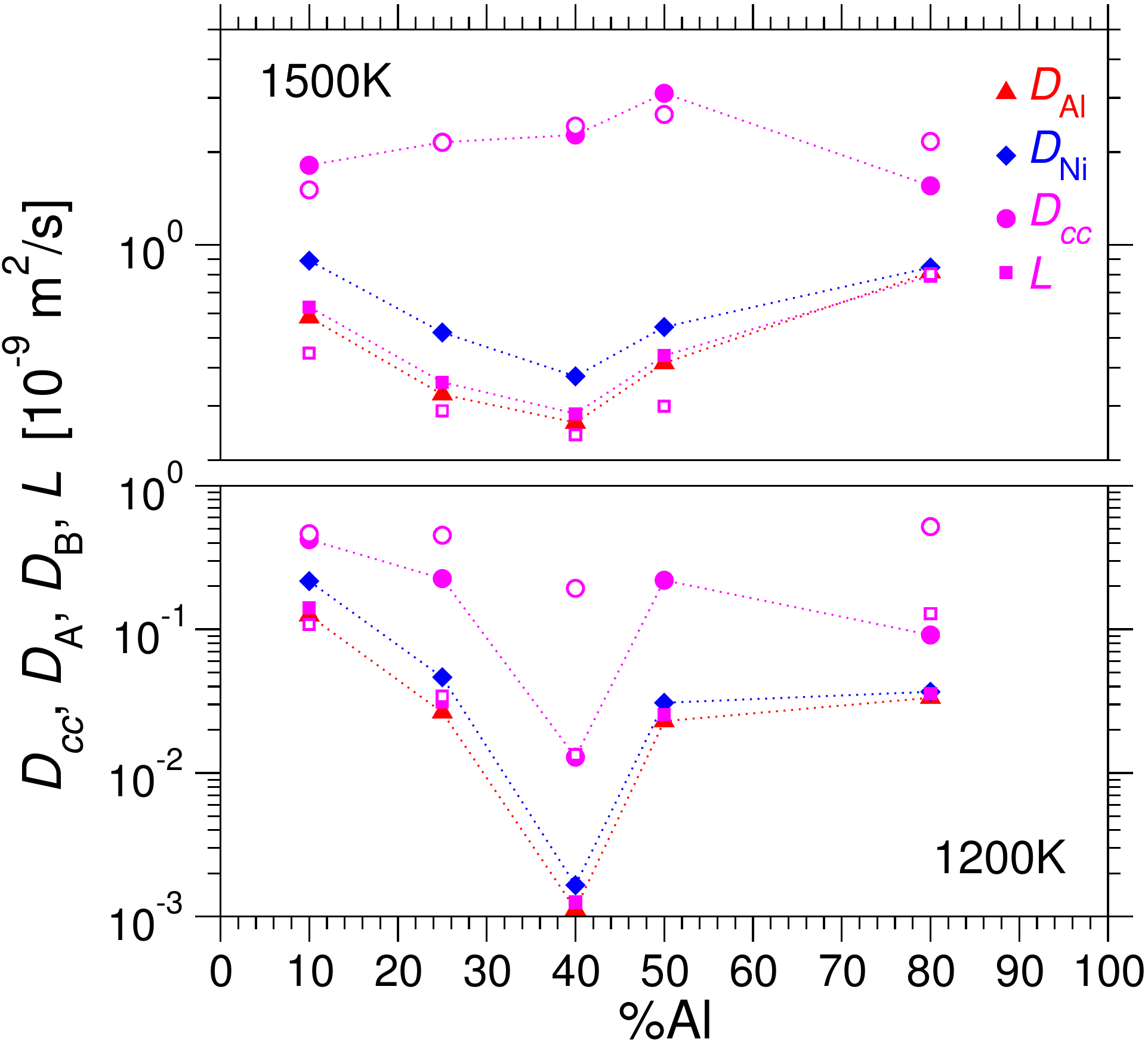}
\caption{\label{alnimctd}
  Self-diffusion coefficients $D_\alpha$, $\alpha=\text{Al}$, $\text{Ni}$,
  interdiffusion coefficient $D_{cc}$, and the corresponding Onsager
  coefficient $L$ as functions of Al-concentration,
  calculated within MCT using the Al-Ni static
  structure factors from the simulation, at $T=1500\,\text{K}$ (upper
  panel) and $T=1200\,\text{K}$ (lower panel).
  MD-simulation results for the interdiffusion coefficient $D_{cc}$ and
  the corresponding Onsager coefficient $L$ are included as open symbols
  for $T=1250\,\text{K}$
  and $T=900\,\text{K}$, scaled by a factor $0.4$.
}
\end{figure}

To verify the MCT description, we show in Fig.~\ref{alnimctd}
the diffusion and Onsager coefficients of Al-Ni as a function of
composition, for fixed temperature. For the MCT calculations,
$T=1500\,\text{K}$ and $T=1200\,\text{K}$ were chosen, still above $T_c$ at
any composition. However,
while for $T=1500\,\text{K}$ at $x\to0$, the distance to the transition is
$\varepsilon=(T_c-T)/T_c\approx-0.67$ (using the standard convention
that states on the liquid side of the glass transition corresdond
to negative distance parameter $\varepsilon$), this distance reduces
to about $\varepsilon\approx-0.25$ around $x\approx0.4$. As a result,
the self-diffusion and Onsager coefficients decrease by about a factor of
$2$ between these two compositions. This explains the minimum also seen
in all kinetic coefficients of the Al-Ni and Zr-Ni simulations discussed
in conjunction with Figs.~\ref{alnids} and \ref{zrnixonet}.
Lowering $T$, as exemplified by the $T=1200\,\text{K}$
result, even smaller absolute values of the distance parameter $\varepsilon$
are reached around $x\approx0.4$, resulting in a stronger drop of
the transport coefficients.

The MCT estimate for the interdiffusion coefficient is obtained from
the Onsager coefficient multiplied with the thermodynamic factor. The latter
is taken directly from the MD-simulated structure factor, and thus is not
by itself an MCT prediction. The result for $D_{cc}$ is shown as the
filled circles in Fig.~\ref{alnimctd}. Since the suppression of the
Onsager coefficient $L$ at $T=1500\,\text{K}$
is about a factor of $2$, but the increase
in the thermodynamic factor $\Phi$ amounts to a factor of $3$ over the
concentration range covered here, cf.\ Fig.~\ref{alnizrniphi}, an
increase of about a factor $1.5$ is observed for $D_{cc}$ over the
same concentration range.

Taking this into account, the qualitative agreement of the MCT results with
those obtained by the MD simulations discussed above, is
quite good. A source of quantitative error is the over-estimation of $T_c$ by
MCT, as discussed above.
It is therefore usual to compare MCT results with those of the MD simulation
at a lower temperature. This is not quite accurate for the present
discussion, since this overemphasizes the variation in the thermodynamic
factor. We therefore have to anticipate that the translation
between $T_\text{MD}$ and $T_\text{MCT}$ where best agreement between simulation
and theory is observed, will not be a simple linear relationship.
To nevertheless demonstrate the level of qualitative agreement, we
include in Fig.~\ref{alnimctd} MD simulation results for some
$T_\text{MD}<T_\text{MCT}$, to adjust for the discrepancy in $T_c$, and
additionaly scaled by empirical factors to account for the mismatch
in thermodynamic factors at the temperatures used in the comparison.
In particular for the higher temperature, $T_\text{MCT}=1500\,\text{K}$,
the agreement is quite good on a qualitative level.

For the lower temperature $T=1200\,\text{K}$, in MCT both the
self-diffusion and Onsager coefficients drop by around two orders of magnitude
upon changing the composition. Even the five-fold increase in the
thermodynamic factor can no longer compensate for this, so that the
interdiffusion coefficient $D_{cc}$ now displays a pronounced minimum
around $x\approx0.4$. Comparing with the
MD simulation results for $T_\text{MD}=900\,\text{K}$ shown in
Fig.~\ref{alnidx}, and added as rescaled quantities in Fig.~\ref{alnimctd},
this minimum is less strong and less narrow in the simulation.
This is expected, since in the region around $x\approx0.4$, our MCT
calculations probe a temperature quite close to $T_c$. Here, the additional
low-temperature transport processes that MCT neglects, are dominant.

For the MCT results,
one even notices a nonmonotonic trend involving a minimum and a subsequent
maximum upon increasing $x$ in $D_{cc}$ at $T=1200$K. This is not
directly confirmed by MD simulations, but highlights what may happen
generically for low temperatures. The concentration of the minimum
in the Onsager coefficient $L$ and the concentration of the maximum in the
thermodynamic factor $\Phi$ do not necessarily coincide, since they
are determined by different physical mechanisms (one kinetic, the other
thermodynamic in origin). $D_{cc}$ as the product of these two
counter-balancing terms may hence even display multiple minima and maxima.

MCT predicts the Darken approximation to hold much better than what is
observed in the simulation. From the theory, for the state points shown
in Fig.~\ref{alnimctd}, $|S-1|\lesssim0.1$ holds. In a sense, the MCT
approximation of representing all relevant correlation functions in
terms of their overlap with density-relaxation modes overemphasizes
the coupling of different dynamical transport mechanisms.
Note also that MCT is constructed to describe the slowing down of
transport processes as $T_c$ is approached. To this end, the theory
is centered around an approximation of the growing dynamical friction
expressed through its memory kernel. In schematic terms, MCT describes
the growth of the friction coefficient for particle motion, $\zeta\to\infty$,
and hence the suppression of diffusivity, $D\sim kT/\zeta\to0$.
In this sense, MCT is a ``slow-mode approximation''.
The Darken equation, Eq.~\eqref{darken}, on the other hand is a typical
``fast-mode approximation''. This conceptual difference between Darken's
approach and the mode-coupling theory can lead to very different
estimates of the interdiffusion coefficient in particular when the
species of the mixture are governed by dynamical processes on very
different time scales (such as in mixtures of very disparate species).
For the binary metallic melts discussed here, this difference is however
not important.

\section{Conclusions}\label{conclusions}

We have studied the concentration dependence of self-diffusion,
$D^s_\alpha$, and interdiffusion, $D_{cc}$, in two
dense metallic melts, Zr-Ni and Al-Ni, by molecular-dynamics simulation
and the mode-coupling theory of the glass transition.
The two systems were chosen as representatives of metallic melts with
very different thermodynamic phase diagrams, involving different eutectica,
and rather different glass-forming ability \cite{Cheng.2011}.
Nevertheless, two generic mixing effects emerge for the mass transport in the
(supercooled) liquid state of these non-demixing systems:
self diffusion becomes slower upon mixing (i.e.,
upon increasing the concentration of the minority species towards $x\approx0.5$)
while thermodynamic driving forces become stronger upon mixing.
The suppression of diffusion kinetics is more pronounced for lower temperatures,
in qualitative agreement with MCT.

Interdiffusion is a combination of a kinetic process,
quantified by Onsager's coefficient $L$, and a thermodynamic driving
force $\Phi$. Consequently, it displays a more subtle mixing scenario.
At high temperatures, the enhancement in thermodynamic driving force
dominates, leading to a maximum in $D_{cc}$ as a function of concentration.
At lower temperatures, the kinetic slowing down
is more dominant, causing a minimum in $D_{cc}$.
This arises although thermodynamic driving forces increase
with decreasing temperature: across the MCT glass transition, thermodynamic
quantities such as $\Phi$ change smoothly, while kinetic coefficients exhibit
pre-cursors of the ideal-glass singularity ($1/L\to\infty$ in this case).
Hence, close to the MCT-$T_c$, the kinetic contributions are dominant.

Since thermodynamic and dynamic contributions to interdiffusion exhibit
maxima respectively minima at slightly different compositions, the
concentration dependence of $D_{cc}$ could in principle be more complicated
and lead to $D_{cc}$-versus-$x$ curves with more than one extremum. Indications
for this are seen in the MD simulation at low temperatures, but the
prediction awaits confirmation.

Note that a similar distinction between dynamic and thermodynamic quantities
as drawn here,
allows to understand non-monotonic behavior in the crystal growth
velocity of glass-forming melts \cite{Wang.2014}, as well as generic
features of the pressure- versus temperature-dependence of the glass
transition \cite{Voigtmann.2008d}.

MCT attributes slow kinetics to the caging of particles by their
neighbors in the dense melt. Diffusion is then governed by
the collective breaking of such cages corresponding to the slow relaxation
of density fluctuations (called $\alpha$ relaxation or
structural relaxation in the glass-transition literature).
Since slow structural relaxation governs all mass-transport processes
close to $T_c$, also the Onsager coefficient $L$ for interdiffusion couples
strongly to the self-diffusion coefficients $D^s_\alpha$ of the individual
species. In consequence, the Darken relation expressing $L$ through the
$D^s_\alpha$ is well fulfilled. For structural-relaxation effects to
dominate, the diffusion processes have to be sufficiently slow, say,
$D<10^{-9}\,\text{m}^2/\text{s}$.
At higher temperatures, short-time correlated binary-collision dynamics as
expressed through the Enskog theory of liquids becomes important;
typically, for a binary mixture of species A and B, this will mean that
A-B cross terms contribute significantly to the dynamics. In this case,
the Darken equation will be violated, expressed by a correction factor
$S$ significantly different from unity.

In the MD simulation, the strongest deviations from Darken's equation
are found indeed in the concentration range around $x\approx0.5$
and at high temperatures.
For the temperatures discussed here, the worst error made by the Darken
equation is around $66\%$. In this region, Darken's assumption is
qualitatively wrong, in the sense that the true Onsager coefficient can be
smaller than either self-diffusion coefficients.
The error of the Darken equation diminishes upon lowering temperature,
at the same time when the dynamic suppression of interdiffusion becomes
dominant over its thermodynamic enhancement.

Qualitative features of the kinetic mass transport processes in dense
metallic melts can often be described by a hard-sphere analogy
\cite{Meyer.2002}, emphasizing their entropic nature. As predicted by
MCT, the slowing down upon mixing is a generic effect for mixtures composed
of similar-sized spheres. A qualitative assessment of the melt's diffusion
dynamics based on experimental data for the density, in terms of an effective
hard-sphere packing fraction and size ratio, indeed gives reasonable
results for Al-Ni, but slightly less so for Zr-Ni (which may indicate
stronger contributions from chemical short-range order \cite{Voigtmann.2008b}).
The strong thermodynamic driving forces seen
in the interdiffusion in metallic melts, on the other hand, are a
non-entropic effect that is only poorly captured by the hard-sphere
analogy.

Our MD simulations intend to capture typical dynamical processes that arise in
metallic melts in addition to the entropic effects caused by strong
excluded-volume interactions. When performing quantitative comparisons,
one has, however, to be aware that the
modeling in terms of effective interaction potentials, such as the
embedded-atom potentials used here, has its limitations. We have based
our discussion on effective potential models that are gauged against
the best available experimental data for the dynamics in the dense melt,
rather than against crystallographic data (which is traditionally the case).
Further experiments, for example on interdiffusion coefficients, will be
needed to judge the quality of the MD models.

The results discussed here for the thermodynamic factor $\Phi$ and the
correction factor $S$, are expected to be generically
valid in mixing systems, where chemical ordering induces quantitative
deviations from the Darken equation, leading to $S\lesssim1$ typically
\cite{Griesche.2009}.
Note that in systems exhibiting liquid--liquid demixing, the behavior
close to this phase transition will be quite
different: at the critical point of this phase transition,
$S$ diverges, but $\Phi$ approaches zero more rapidly
and the interdiffusion coefficient vanishes
\cite{Manning.1961,Das.2006}.
For melts where precursors of the vicinity to such a phase transition
are important, additional generic mixing phenomena (connected to the fact
that $\Phi<1$, as, e.g., in Ag-Cu) will arise that are
not discussed in the present paper.

\begin{acknowledgments}
We thank D.~Holland-Moritz and J.~Brillo for discussion.
Part of this work was supported by the Helmholtz-Gemeinschaft,
Helmholtz-Hochschul-Nachwuchsgruppe VH-NG 406.
\end{acknowledgments}

\bibliography{lit}

\begin{thebibliography}{67}%
\makeatletter
\providecommand \@ifxundefined [1]{%
 \@ifx{#1\undefined}
}%
\providecommand \@ifnum [1]{%
 \ifnum #1\expandafter \@firstoftwo
 \else \expandafter \@secondoftwo
 \fi
}%
\providecommand \@ifx [1]{%
 \ifx #1\expandafter \@firstoftwo
 \else \expandafter \@secondoftwo
 \fi
}%
\providecommand \natexlab [1]{#1}%
\providecommand \enquote  [1]{``#1''}%
\providecommand \bibnamefont  [1]{#1}%
\providecommand \bibfnamefont [1]{#1}%
\providecommand \citenamefont [1]{#1}%
\providecommand \href@noop [0]{\@secondoftwo}%
\providecommand \href [0]{\begingroup \@sanitize@url \@href}%
\providecommand \@href[1]{\@@startlink{#1}\@@href}%
\providecommand \@@href[1]{\endgroup#1\@@endlink}%
\providecommand \@sanitize@url [0]{\catcode `\\12\catcode `\$12\catcode
  `\&12\catcode `\#12\catcode `\^12\catcode `\_12\catcode `\%12\relax}%
\providecommand \@@startlink[1]{}%
\providecommand \@@endlink[0]{}%
\providecommand \url  [0]{\begingroup\@sanitize@url \@url }%
\providecommand \@url [1]{\endgroup\@href {#1}{\urlprefix }}%
\providecommand \urlprefix  [0]{URL }%
\providecommand \Eprint [0]{\href }%
\providecommand \doibase [0]{http://dx.doi.org/}%
\providecommand \selectlanguage [0]{\@gobble}%
\providecommand \bibinfo  [0]{\@secondoftwo}%
\providecommand \bibfield  [0]{\@secondoftwo}%
\providecommand \translation [1]{[#1]}%
\providecommand \BibitemOpen [0]{}%
\providecommand \bibitemStop [0]{}%
\providecommand \bibitemNoStop [0]{.\EOS\space}%
\providecommand \EOS [0]{\spacefactor3000\relax}%
\providecommand \BibitemShut  [1]{\csname bibitem#1\endcsname}%
\let\auto@bib@innerbib\@empty
\bibitem [{\citenamefont {Faupel}\ \emph {et~al.}(2003)\citenamefont {Faupel},
  \citenamefont {Frank}, \citenamefont {Macht}, \citenamefont {Mehrer},
  \citenamefont {Naundort}, \citenamefont {R\"atzke}, \citenamefont {Schober},
  \citenamefont {Sharma},\ and\ \citenamefont {Teichler}}]{Faupel.2003}%
  \BibitemOpen
  \bibfield  {author} {\bibinfo {author} {\bibfnamefont {F.}~\bibnamefont
  {Faupel}}, \bibinfo {author} {\bibfnamefont {W.}~\bibnamefont {Frank}},
  \bibinfo {author} {\bibfnamefont {M.-P.}\ \bibnamefont {Macht}}, \bibinfo
  {author} {\bibfnamefont {H.}~\bibnamefont {Mehrer}}, \bibinfo {author}
  {\bibfnamefont {V.}~\bibnamefont {Naundort}}, \bibinfo {author}
  {\bibfnamefont {K.}~\bibnamefont {R\"atzke}}, \bibinfo {author}
  {\bibfnamefont {H.~R.}\ \bibnamefont {Schober}}, \bibinfo {author}
  {\bibfnamefont {S.~K.}\ \bibnamefont {Sharma}}, \ and\ \bibinfo {author}
  {\bibfnamefont {H.}~\bibnamefont {Teichler}},\ }\href@noop {} {\bibfield
  {journal} {\bibinfo  {journal} {Rev. Mod. Phys.}\ }\textbf {\bibinfo {volume}
  {75}},\ \bibinfo {pages} {237} (\bibinfo {year} {2003})}\BibitemShut
  {NoStop}%
\bibitem [{\citenamefont {Kasperovich}\ \emph {et~al.}(2010)\citenamefont
  {Kasperovich}, \citenamefont {Meyer},\ and\ \citenamefont
  {Ratke}}]{Kasperovich.2010}%
  \BibitemOpen
  \bibfield  {author} {\bibinfo {author} {\bibfnamefont {G.}~\bibnamefont
  {Kasperovich}}, \bibinfo {author} {\bibfnamefont {A.}~\bibnamefont {Meyer}},
  \ and\ \bibinfo {author} {\bibfnamefont {L.}~\bibnamefont {Ratke}},\
  }\href@noop {} {\bibfield  {journal} {\bibinfo  {journal} {Int. Foundry
  Res.}\ }\textbf {\bibinfo {volume} {62(4)}},\ \bibinfo {pages} {8} (\bibinfo
  {year} {2010})}\BibitemShut {NoStop}%
\bibitem [{\citenamefont {Boon}\ and\ \citenamefont {Yip}(1980)}]{Boon}%
  \BibitemOpen
  \bibfield  {author} {\bibinfo {author} {\bibfnamefont {J.-P.}\ \bibnamefont
  {Boon}}\ and\ \bibinfo {author} {\bibfnamefont {S.}~\bibnamefont {Yip}},\
  }\href@noop {} {\emph {\bibinfo {title} {Molecular Hydrodynamics}}}\
  (\bibinfo  {publisher} {McGraw Hill},\ \bibinfo {address} {New York},\
  \bibinfo {year} {1980})\BibitemShut {NoStop}%
\bibitem [{\citenamefont {G\"otze}(2009)}]{Goetze.2009}%
  \BibitemOpen
  \bibfield  {author} {\bibinfo {author} {\bibfnamefont {W.}~\bibnamefont
  {G\"otze}},\ }\href@noop {} {\emph {\bibinfo {title} {Complex Dynamics of
  Glass-Forming Liquids}}}\ (\bibinfo  {publisher} {Oxford University Press},\
  \bibinfo {year} {2009})\BibitemShut {NoStop}%
\bibitem [{\citenamefont {Meyer}(2002)}]{Meyer.2002}%
  \BibitemOpen
  \bibfield  {author} {\bibinfo {author} {\bibfnamefont {A.}~\bibnamefont
  {Meyer}},\ }\href@noop {} {\bibfield  {journal} {\bibinfo  {journal}
  {Phys.~Rev.~B}\ }\textbf {\bibinfo {volume} {66}},\ \bibinfo {pages} {134205}
  (\bibinfo {year} {2002})}\BibitemShut {NoStop}%
\bibitem [{\citenamefont {Brillo}\ \emph {et~al.}(2011)\citenamefont {Brillo},
  \citenamefont {Pommrich},\ and\ \citenamefont {Meyer}}]{Brillo.2011}%
  \BibitemOpen
  \bibfield  {author} {\bibinfo {author} {\bibfnamefont {J.}~\bibnamefont
  {Brillo}}, \bibinfo {author} {\bibfnamefont {A.~I.}\ \bibnamefont
  {Pommrich}}, \ and\ \bibinfo {author} {\bibfnamefont {A.}~\bibnamefont
  {Meyer}},\ }\href@noop {} {\bibfield  {journal} {\bibinfo  {journal} {Phys.
  Rev. Lett.}\ }\textbf {\bibinfo {volume} {107}},\ \bibinfo {pages} {165902}
  (\bibinfo {year} {2011})}\BibitemShut {NoStop}%
\bibitem [{\citenamefont {Shimoji}\ and\ \citenamefont
  {Itami}(1986)}]{Shimoji.1986}%
  \BibitemOpen
  \bibfield  {author} {\bibinfo {author} {\bibfnamefont {M.}~\bibnamefont
  {Shimoji}}\ and\ \bibinfo {author} {\bibfnamefont {T.}~\bibnamefont
  {Itami}},\ }\href@noop {} {\bibfield  {journal} {\bibinfo  {journal} {Defect
  and Diffusion Forum}\ }\textbf {\bibinfo {volume} {43}},\ \bibinfo {pages}
  {154} (\bibinfo {year} {1986})}\BibitemShut {NoStop}%
\bibitem [{\citenamefont {Griesche}\ \emph {et~al.}(2004)\citenamefont
  {Griesche}, \citenamefont {Macht}, \citenamefont {Garandet},\ and\
  \citenamefont {Frohberg}}]{Griesche.2004}%
  \BibitemOpen
  \bibfield  {author} {\bibinfo {author} {\bibfnamefont {A.}~\bibnamefont
  {Griesche}}, \bibinfo {author} {\bibfnamefont {M.-P.}\ \bibnamefont {Macht}},
  \bibinfo {author} {\bibfnamefont {J.-P.}\ \bibnamefont {Garandet}}, \ and\
  \bibinfo {author} {\bibfnamefont {G.}~\bibnamefont {Frohberg}},\ }\href@noop
  {} {\bibfield  {journal} {\bibinfo  {journal} {J. Non-Cryst. Solids}\
  }\textbf {\bibinfo {volume} {336}},\ \bibinfo {pages} {173} (\bibinfo {year}
  {2004})}\BibitemShut {NoStop}%
\bibitem [{\citenamefont {Zhang}\ \emph {et~al.}(2010)\citenamefont {Zhang},
  \citenamefont {Griesche},\ and\ \citenamefont {Meyer}}]{Zhang.2010}%
  \BibitemOpen
  \bibfield  {author} {\bibinfo {author} {\bibfnamefont {B.}~\bibnamefont
  {Zhang}}, \bibinfo {author} {\bibfnamefont {A.}~\bibnamefont {Griesche}}, \
  and\ \bibinfo {author} {\bibfnamefont {A.}~\bibnamefont {Meyer}},\
  }\href@noop {} {\bibfield  {journal} {\bibinfo  {journal} {Phys. Rev. Lett.}\
  }\textbf {\bibinfo {volume} {104}},\ \bibinfo {pages} {035902} (\bibinfo
  {year} {2010})}\BibitemShut {NoStop}%
\bibitem [{\citenamefont {Kargl}\ \emph {et~al.}(2011)\citenamefont {Kargl},
  \citenamefont {Engelhardt}, \citenamefont {Yang}, \citenamefont {Weis},
  \citenamefont {Schmakat}, \citenamefont {Schillinger}, \citenamefont
  {Griesche},\ and\ \citenamefont {Meyer}}]{Kargl.2011}%
  \BibitemOpen
  \bibfield  {author} {\bibinfo {author} {\bibfnamefont {F.}~\bibnamefont
  {Kargl}}, \bibinfo {author} {\bibfnamefont {M.}~\bibnamefont {Engelhardt}},
  \bibinfo {author} {\bibfnamefont {F.}~\bibnamefont {Yang}}, \bibinfo {author}
  {\bibfnamefont {H.}~\bibnamefont {Weis}}, \bibinfo {author} {\bibfnamefont
  {P.}~\bibnamefont {Schmakat}}, \bibinfo {author} {\bibfnamefont
  {B.}~\bibnamefont {Schillinger}}, \bibinfo {author} {\bibfnamefont
  {A.}~\bibnamefont {Griesche}}, \ and\ \bibinfo {author} {\bibfnamefont
  {A.}~\bibnamefont {Meyer}},\ }\href@noop {} {\bibfield  {journal} {\bibinfo
  {journal} {J. Phys.: Condens. Matter}\ }\textbf {\bibinfo {volume} {23}},\
  \bibinfo {pages} {254201} (\bibinfo {year} {2011})}\BibitemShut {NoStop}%
\bibitem [{\citenamefont {Kargl}\ \emph {et~al.}(2013)\citenamefont {Kargl},
  \citenamefont {Sondermann}, \citenamefont {Weis},\ and\ \citenamefont
  {Meyer}}]{Kargl.2013}%
  \BibitemOpen
  \bibfield  {author} {\bibinfo {author} {\bibfnamefont {F.}~\bibnamefont
  {Kargl}}, \bibinfo {author} {\bibfnamefont {E.}~\bibnamefont {Sondermann}},
  \bibinfo {author} {\bibfnamefont {H.}~\bibnamefont {Weis}}, \ and\ \bibinfo
  {author} {\bibfnamefont {A.}~\bibnamefont {Meyer}},\ }\href@noop {}
  {\bibfield  {journal} {\bibinfo  {journal} {High Temp. High Press.}\ }\textbf
  {\bibinfo {volume} {42}},\ \bibinfo {pages} {3} (\bibinfo {year}
  {2013})}\BibitemShut {NoStop}%
\bibitem [{\citenamefont {Garandet}\ \emph {et~al.}(1995)\citenamefont
  {Garandet}, \citenamefont {Barat},\ and\ \citenamefont
  {Duffar}}]{Garandet.1995}%
  \BibitemOpen
  \bibfield  {author} {\bibinfo {author} {\bibfnamefont {J.~P.}\ \bibnamefont
  {Garandet}}, \bibinfo {author} {\bibfnamefont {C.}~\bibnamefont {Barat}}, \
  and\ \bibinfo {author} {\bibfnamefont {T.}~\bibnamefont {Duffar}},\
  }\href@noop {} {\bibfield  {journal} {\bibinfo  {journal} {Int. J. Heat Mass
  Transf.}\ }\textbf {\bibinfo {volume} {38}},\ \bibinfo {pages} {2169}
  (\bibinfo {year} {1995})}\BibitemShut {NoStop}%
\bibitem [{\citenamefont {Barat}\ and\ \citenamefont
  {Garandet}(1996)}]{Barat.1996}%
  \BibitemOpen
  \bibfield  {author} {\bibinfo {author} {\bibfnamefont {C.}~\bibnamefont
  {Barat}}\ and\ \bibinfo {author} {\bibfnamefont {J.~P.}\ \bibnamefont
  {Garandet}},\ }\href@noop {} {\bibfield  {journal} {\bibinfo  {journal} {Int.
  J. Heat Mass Transf.}\ }\textbf {\bibinfo {volume} {39}},\ \bibinfo {pages}
  {2177} (\bibinfo {year} {1996})}\BibitemShut {NoStop}%
\bibitem [{\citenamefont {Sutherland}(1905)}]{Sutherland.1905}%
  \BibitemOpen
  \bibfield  {author} {\bibinfo {author} {\bibfnamefont {W.}~\bibnamefont
  {Sutherland}},\ }\href@noop {} {\bibfield  {journal} {\bibinfo  {journal}
  {Phil. Mag. Ser. 6}\ }\textbf {\bibinfo {volume} {9}},\ \bibinfo {pages}
  {781} (\bibinfo {year} {1905})}\BibitemShut {NoStop}%
\bibitem [{\citenamefont {{De Michele}}\ and\ \citenamefont
  {Leporini}(2001)}]{DeMichele.2001}%
  \BibitemOpen
  \bibfield  {author} {\bibinfo {author} {\bibfnamefont {C.}~\bibnamefont {{De
  Michele}}}\ and\ \bibinfo {author} {\bibfnamefont {D.}~\bibnamefont
  {Leporini}},\ }\href@noop {} {\bibfield  {journal} {\bibinfo  {journal}
  {Phys. Rev. E}\ }\textbf {\bibinfo {volume} {63}},\ \bibinfo {pages} {036701}
  (\bibinfo {year} {2001})}\BibitemShut {NoStop}%
\bibitem [{\citenamefont {Fujara}\ \emph {et~al.}(1992)\citenamefont {Fujara},
  \citenamefont {Geil}, \citenamefont {Sillescu},\ and\ \citenamefont
  {Fleischer}}]{Fujara.1992}%
  \BibitemOpen
  \bibfield  {author} {\bibinfo {author} {\bibfnamefont {F.}~\bibnamefont
  {Fujara}}, \bibinfo {author} {\bibfnamefont {B.}~\bibnamefont {Geil}},
  \bibinfo {author} {\bibfnamefont {H.}~\bibnamefont {Sillescu}}, \ and\
  \bibinfo {author} {\bibfnamefont {G.}~\bibnamefont {Fleischer}},\ }\href@noop
  {} {\bibfield  {journal} {\bibinfo  {journal} {Z. Phys. B}\ }\textbf
  {\bibinfo {volume} {88}},\ \bibinfo {pages} {195} (\bibinfo {year}
  {1992})}\BibitemShut {NoStop}%
\bibitem [{\citenamefont {Ediger}(2000)}]{Ediger.2000}%
  \BibitemOpen
  \bibfield  {author} {\bibinfo {author} {\bibfnamefont {M.~D.}\ \bibnamefont
  {Ediger}},\ }\href@noop {} {\bibfield  {journal} {\bibinfo  {journal} {Annu.
  Rev. Phys. Chem.}\ }\textbf {\bibinfo {volume} {51}},\ \bibinfo {pages} {99}
  (\bibinfo {year} {2000})}\BibitemShut {NoStop}%
\bibitem [{\citenamefont {Zwanzig}\ and\ \citenamefont
  {Harrison}(1985)}]{Zwanzig.1985}%
  \BibitemOpen
  \bibfield  {author} {\bibinfo {author} {\bibfnamefont {R.}~\bibnamefont
  {Zwanzig}}\ and\ \bibinfo {author} {\bibfnamefont {A.~K.}\ \bibnamefont
  {Harrison}},\ }\href@noop {} {\bibfield  {journal} {\bibinfo  {journal} {J.
  Chem. Phys.}\ }\textbf {\bibinfo {volume} {83}},\ \bibinfo {pages} {5861}
  (\bibinfo {year} {1985})}\BibitemShut {NoStop}%
\bibitem [{\citenamefont {Puertas}\ and\ \citenamefont
  {Voigtmann}(2014)}]{Puertas.2014}%
  \BibitemOpen
  \bibfield  {author} {\bibinfo {author} {\bibfnamefont {A.~M.}\ \bibnamefont
  {Puertas}}\ and\ \bibinfo {author} {\bibfnamefont {{\relax Th}.}~\bibnamefont
  {Voigtmann}},\ }\href@noop {} {\enquote {\bibinfo {title} {Microrheology of
  colloidal systems},}\ } (\bibinfo {year} {2014}),\ \bibinfo {note}
  {{s}ubmitted}\BibitemShut {NoStop}%
\bibitem [{\citenamefont {Darken}(1948)}]{Darken.1948}%
  \BibitemOpen
  \bibfield  {author} {\bibinfo {author} {\bibfnamefont {L.~S.}\ \bibnamefont
  {Darken}},\ }\href@noop {} {\bibfield  {journal} {\bibinfo  {journal} {Trans.
  AIME}\ }\textbf {\bibinfo {volume} {175}},\ \bibinfo {pages} {184} (\bibinfo
  {year} {1948})}\BibitemShut {NoStop}%
\bibitem [{\citenamefont {Horbach}\ \emph {et~al.}(2007)\citenamefont
  {Horbach}, \citenamefont {Das}, \citenamefont {Griesche}, \citenamefont
  {Macht}, \citenamefont {Froh\-berg},\ and\ \citenamefont
  {Meyer}}]{Horbach.2007}%
  \BibitemOpen
  \bibfield  {author} {\bibinfo {author} {\bibfnamefont {J.}~\bibnamefont
  {Horbach}}, \bibinfo {author} {\bibfnamefont {S.~K.}\ \bibnamefont {Das}},
  \bibinfo {author} {\bibfnamefont {A.}~\bibnamefont {Griesche}}, \bibinfo
  {author} {\bibfnamefont {M.-P.}\ \bibnamefont {Macht}}, \bibinfo {author}
  {\bibfnamefont {G.}~\bibnamefont {Froh\-berg}}, \ and\ \bibinfo {author}
  {\bibfnamefont {A.}~\bibnamefont {Meyer}},\ }\href@noop {} {\bibfield
  {journal} {\bibinfo  {journal} {Phys.~Rev.~B}\ }\textbf {\bibinfo {volume}
  {75}},\ \bibinfo {pages} {174304} (\bibinfo {year} {2007})}\BibitemShut
  {NoStop}%
\bibitem [{\citenamefont {Hansen}\ and\ \citenamefont
  {McDonald}(1986)}]{Hansen}%
  \BibitemOpen
  \bibfield  {author} {\bibinfo {author} {\bibfnamefont {J.-P.}\ \bibnamefont
  {Hansen}}\ and\ \bibinfo {author} {\bibfnamefont {I.~R.}\ \bibnamefont
  {McDonald}},\ }\href@noop {} {\emph {\bibinfo {title} {Theory of Simple
  Liquids}}}\ (\bibinfo  {publisher} {Academic Press},\ \bibinfo {address}
  {London},\ \bibinfo {year} {1986})\BibitemShut {NoStop}%
\bibitem [{\citenamefont {Manning}(1961)}]{Manning.1961}%
  \BibitemOpen
  \bibfield  {author} {\bibinfo {author} {\bibfnamefont {J.~R.}\ \bibnamefont
  {Manning}},\ }\href@noop {} {\bibfield  {journal} {\bibinfo  {journal} {Phys.
  Rev.}\ }\textbf {\bibinfo {volume} {124}},\ \bibinfo {pages} {470} (\bibinfo
  {year} {1961})}\BibitemShut {NoStop}%
\bibitem [{\citenamefont {Ehmler}\ \emph {et~al.}(1998)\citenamefont {Ehmler},
  \citenamefont {Heesemann}, \citenamefont {R\"atzke}, \citenamefont {Faupel},\
  and\ \citenamefont {Geyer}}]{Ehmler.1998}%
  \BibitemOpen
  \bibfield  {author} {\bibinfo {author} {\bibfnamefont {H.}~\bibnamefont
  {Ehmler}}, \bibinfo {author} {\bibfnamefont {A.}~\bibnamefont {Heesemann}},
  \bibinfo {author} {\bibfnamefont {K.}~\bibnamefont {R\"atzke}}, \bibinfo
  {author} {\bibfnamefont {F.}~\bibnamefont {Faupel}}, \ and\ \bibinfo {author}
  {\bibfnamefont {U.}~\bibnamefont {Geyer}},\ }\href@noop {} {\bibfield
  {journal} {\bibinfo  {journal} {Phys. Rev. Lett.}\ }\textbf {\bibinfo
  {volume} {80}},\ \bibinfo {pages} {4919} (\bibinfo {year}
  {1998})}\BibitemShut {NoStop}%
\bibitem [{\citenamefont {Z\"ollmer}\ \emph {et~al.}(2003)\citenamefont
  {Z\"ollmer}, \citenamefont {R\"atzke}, \citenamefont {Faupel},\ and\
  \citenamefont {Meyer}}]{Zoellmer.2003}%
  \BibitemOpen
  \bibfield  {author} {\bibinfo {author} {\bibfnamefont {V.}~\bibnamefont
  {Z\"ollmer}}, \bibinfo {author} {\bibfnamefont {K.}~\bibnamefont {R\"atzke}},
  \bibinfo {author} {\bibfnamefont {F.}~\bibnamefont {Faupel}}, \ and\ \bibinfo
  {author} {\bibfnamefont {A.}~\bibnamefont {Meyer}},\ }\href@noop {}
  {\bibfield  {journal} {\bibinfo  {journal} {Phys. Rev. Lett.}\ }\textbf
  {\bibinfo {volume} {90}},\ \bibinfo {pages} {195502} (\bibinfo {year}
  {2003})}\BibitemShut {NoStop}%
\bibitem [{\citenamefont {Bartsch}\ \emph {et~al.}(2006)\citenamefont
  {Bartsch}, \citenamefont {R\"atzke}, \citenamefont {Faupel},\ and\
  \citenamefont {Meyer}}]{Bartsch.2006}%
  \BibitemOpen
  \bibfield  {author} {\bibinfo {author} {\bibfnamefont {A.}~\bibnamefont
  {Bartsch}}, \bibinfo {author} {\bibfnamefont {K.}~\bibnamefont {R\"atzke}},
  \bibinfo {author} {\bibfnamefont {F.}~\bibnamefont {Faupel}}, \ and\ \bibinfo
  {author} {\bibfnamefont {A.}~\bibnamefont {Meyer}},\ }\href@noop {}
  {\bibfield  {journal} {\bibinfo  {journal} {Appl. Phys. Lett.}\ }\textbf
  {\bibinfo {volume} {89}},\ \bibinfo {pages} {121917} (\bibinfo {year}
  {2006})}\BibitemShut {NoStop}%
\bibitem [{\citenamefont {Bartsch}\ \emph {et~al.}(2010)\citenamefont
  {Bartsch}, \citenamefont {R\"atzke}, \citenamefont {Meyer},\ and\
  \citenamefont {Faupel}}]{Bartsch.2010}%
  \BibitemOpen
  \bibfield  {author} {\bibinfo {author} {\bibfnamefont {A.}~\bibnamefont
  {Bartsch}}, \bibinfo {author} {\bibfnamefont {K.}~\bibnamefont {R\"atzke}},
  \bibinfo {author} {\bibfnamefont {A.}~\bibnamefont {Meyer}}, \ and\ \bibinfo
  {author} {\bibfnamefont {F.}~\bibnamefont {Faupel}},\ }\href@noop {}
  {\bibfield  {journal} {\bibinfo  {journal} {Phys. Rev. Lett.}\ }\textbf
  {\bibinfo {volume} {104}},\ \bibinfo {pages} {195901} (\bibinfo {year}
  {2010})}\BibitemShut {NoStop}%
\bibitem [{\citenamefont {Baxter}(1970)}]{Baxter.1970}%
  \BibitemOpen
  \bibfield  {author} {\bibinfo {author} {\bibfnamefont {R.~J.}\ \bibnamefont
  {Baxter}},\ }\href@noop {} {\bibfield  {journal} {\bibinfo  {journal} {J.
  Chem. Phys.}\ }\textbf {\bibinfo {volume} {52}},\ \bibinfo {pages} {4559}
  (\bibinfo {year} {1970})}\BibitemShut {NoStop}%
\bibitem [{\citenamefont {Barrat}\ and\ \citenamefont
  {Latz}(1990)}]{Barrat.1990}%
  \BibitemOpen
  \bibfield  {author} {\bibinfo {author} {\bibfnamefont {J.-L.}\ \bibnamefont
  {Barrat}}\ and\ \bibinfo {author} {\bibfnamefont {A.}~\bibnamefont {Latz}},\
  }\href@noop {} {\bibfield  {journal} {\bibinfo  {journal} {J. Phys.: Condens.
  Matter}\ }\textbf {\bibinfo {volume} {2}},\ \bibinfo {pages} {4289} (\bibinfo
  {year} {1990})}\BibitemShut {NoStop}%
\bibitem [{\citenamefont {G\"otze}\ and\ \citenamefont
  {Voigtmann}(2003)}]{Goetze.2003}%
  \BibitemOpen
  \bibfield  {author} {\bibinfo {author} {\bibfnamefont {W.}~\bibnamefont
  {G\"otze}}\ and\ \bibinfo {author} {\bibfnamefont {{\relax Th}.}~\bibnamefont
  {Voigtmann}},\ }\href@noop {} {\bibfield  {journal} {\bibinfo  {journal}
  {Phys.~Rev.~E}\ }\textbf {\bibinfo {volume} {67}},\ \bibinfo {pages} {021502}
  (\bibinfo {year} {2003})}\BibitemShut {NoStop}%
\bibitem [{\citenamefont {Foffi}\ \emph {et~al.}(2003)\citenamefont {Foffi},
  \citenamefont {G\"otze}, \citenamefont {Sciortino}, \citenamefont
  {Tartaglia},\ and\ \citenamefont {Voigtmann}}]{Foffi.2003}%
  \BibitemOpen
  \bibfield  {author} {\bibinfo {author} {\bibfnamefont {G.}~\bibnamefont
  {Foffi}}, \bibinfo {author} {\bibfnamefont {W.}~\bibnamefont {G\"otze}},
  \bibinfo {author} {\bibfnamefont {F.}~\bibnamefont {Sciortino}}, \bibinfo
  {author} {\bibfnamefont {P.}~\bibnamefont {Tartaglia}}, \ and\ \bibinfo
  {author} {\bibfnamefont {{\relax Th}.}~\bibnamefont {Voigtmann}},\
  }\href@noop {} {\bibfield  {journal} {\bibinfo  {journal} {Phys.~Rev.~Lett.}\
  }\textbf {\bibinfo {volume} {91}},\ \bibinfo {pages} {085701} (\bibinfo
  {year} {2003})}\BibitemShut {NoStop}%
\bibitem [{\citenamefont {Foffi}\ \emph {et~al.}(2004)\citenamefont {Foffi},
  \citenamefont {G\"otze}, \citenamefont {Sciortino}, \citenamefont
  {Tartaglia},\ and\ \citenamefont {Voigtmann}}]{Foffi.2004}%
  \BibitemOpen
  \bibfield  {author} {\bibinfo {author} {\bibfnamefont {G.}~\bibnamefont
  {Foffi}}, \bibinfo {author} {\bibfnamefont {W.}~\bibnamefont {G\"otze}},
  \bibinfo {author} {\bibfnamefont {F.}~\bibnamefont {Sciortino}}, \bibinfo
  {author} {\bibfnamefont {P.}~\bibnamefont {Tartaglia}}, \ and\ \bibinfo
  {author} {\bibfnamefont {{\relax Th}.}~\bibnamefont {Voigtmann}},\
  }\href@noop {} {\bibfield  {journal} {\bibinfo  {journal} {Phys.~Rev.~E}\
  }\textbf {\bibinfo {volume} {69}},\ \bibinfo {pages} {011505} (\bibinfo
  {year} {2004})}\BibitemShut {NoStop}%
\bibitem [{\citenamefont {Voigtmann}\ \emph {et~al.}(2008)\citenamefont
  {Voigtmann}, \citenamefont {Meyer}, \citenamefont {Holland-Moritz},
  \citenamefont {St\"uber}, \citenamefont {Hansen},\ and\ \citenamefont
  {Unruh}}]{Voigtmann.2008b}%
  \BibitemOpen
  \bibfield  {author} {\bibinfo {author} {\bibfnamefont {{\relax
  Th}.}~\bibnamefont {Voigtmann}}, \bibinfo {author} {\bibfnamefont
  {A.}~\bibnamefont {Meyer}}, \bibinfo {author} {\bibfnamefont
  {D.}~\bibnamefont {Holland-Moritz}}, \bibinfo {author} {\bibfnamefont
  {S.}~\bibnamefont {St\"uber}}, \bibinfo {author} {\bibfnamefont
  {T.}~\bibnamefont {Hansen}}, \ and\ \bibinfo {author} {\bibfnamefont
  {T.}~\bibnamefont {Unruh}},\ }\href@noop {} {\bibfield  {journal} {\bibinfo
  {journal} {EPL}\ }\textbf {\bibinfo {volume} {82}},\ \bibinfo {pages} {66001}
  (\bibinfo {year} {2008})}\BibitemShut {NoStop}%
\bibitem [{\citenamefont {Das}\ \emph {et~al.}(2005)\citenamefont {Das},
  \citenamefont {Horbach}, \citenamefont {Koza}, \citenamefont {{Mavila
  Chatoth}},\ and\ \citenamefont {Meyer}}]{Das.2005}%
  \BibitemOpen
  \bibfield  {author} {\bibinfo {author} {\bibfnamefont {S.~K.}\ \bibnamefont
  {Das}}, \bibinfo {author} {\bibfnamefont {J.}~\bibnamefont {Horbach}},
  \bibinfo {author} {\bibfnamefont {M.~M.}\ \bibnamefont {Koza}}, \bibinfo
  {author} {\bibfnamefont {S.}~\bibnamefont {{Mavila Chatoth}}}, \ and\
  \bibinfo {author} {\bibfnamefont {A.}~\bibnamefont {Meyer}},\ }\href@noop {}
  {\bibfield  {journal} {\bibinfo  {journal} {Appl. Phys. Lett.}\ }\textbf
  {\bibinfo {volume} {86}},\ \bibinfo {pages} {011918} (\bibinfo {year}
  {2005})}\BibitemShut {NoStop}%
\bibitem [{\citenamefont {Griesche}\ \emph {et~al.}(2009)\citenamefont
  {Griesche}, \citenamefont {Zhang}, \citenamefont {Horbach},\ and\
  \citenamefont {Meyer}}]{Griesche.2009}%
  \BibitemOpen
  \bibfield  {author} {\bibinfo {author} {\bibfnamefont {A.}~\bibnamefont
  {Griesche}}, \bibinfo {author} {\bibfnamefont {B.}~\bibnamefont {Zhang}},
  \bibinfo {author} {\bibfnamefont {J.}~\bibnamefont {Horbach}}, \ and\
  \bibinfo {author} {\bibfnamefont {A.}~\bibnamefont {Meyer}},\ }\href@noop {}
  {\bibfield  {journal} {\bibinfo  {journal} {Def. Diff. Forum}\ }\textbf
  {\bibinfo {volume} {289--292}},\ \bibinfo {pages} {705} (\bibinfo {year}
  {2009})}\BibitemShut {NoStop}%
\bibitem [{\citenamefont {St\"uber}\ \emph {et~al.}(2010)\citenamefont
  {St\"uber}, \citenamefont {Holland-Moritz}, \citenamefont {Unruh},\ and\
  \citenamefont {Meyer}}]{Stueber.2010}%
  \BibitemOpen
  \bibfield  {author} {\bibinfo {author} {\bibfnamefont {S.}~\bibnamefont
  {St\"uber}}, \bibinfo {author} {\bibfnamefont {D.}~\bibnamefont
  {Holland-Moritz}}, \bibinfo {author} {\bibfnamefont {T.}~\bibnamefont
  {Unruh}}, \ and\ \bibinfo {author} {\bibfnamefont {A.}~\bibnamefont
  {Meyer}},\ }\href@noop {} {\bibfield  {journal} {\bibinfo  {journal} {Phys.
  Rev. B}\ }\textbf {\bibinfo {volume} {81}},\ \bibinfo {pages} {024204}
  (\bibinfo {year} {2010})}\BibitemShut {NoStop}%
\bibitem [{\citenamefont {Mishin}\ \emph {et~al.}(2002)\citenamefont {Mishin},
  \citenamefont {Mehl},\ and\ \citenamefont
  {Papaconstantopoulos}}]{Mishin.2002}%
  \BibitemOpen
  \bibfield  {author} {\bibinfo {author} {\bibfnamefont {Y.}~\bibnamefont
  {Mishin}}, \bibinfo {author} {\bibfnamefont {M.~J.}\ \bibnamefont {Mehl}}, \
  and\ \bibinfo {author} {\bibfnamefont {D.~A.}\ \bibnamefont
  {Papaconstantopoulos}},\ }\href@noop {} {\bibfield  {journal} {\bibinfo
  {journal} {Phys. Rev. B}\ }\textbf {\bibinfo {volume} {65}},\ \bibinfo
  {pages} {224114} (\bibinfo {year} {2002})}\BibitemShut {NoStop}%
\bibitem [{\citenamefont {Kerrache}\ \emph {et~al.}(2008)\citenamefont
  {Kerrache}, \citenamefont {Horbach},\ and\ \citenamefont
  {Binder}}]{Kerrache.2008}%
  \BibitemOpen
  \bibfield  {author} {\bibinfo {author} {\bibfnamefont {A.}~\bibnamefont
  {Kerrache}}, \bibinfo {author} {\bibfnamefont {J.}~\bibnamefont {Horbach}}, \
  and\ \bibinfo {author} {\bibfnamefont {K.}~\bibnamefont {Binder}},\
  }\href@noop {} {\bibfield  {journal} {\bibinfo  {journal} {EPL}\ }\textbf
  {\bibinfo {volume} {81}},\ \bibinfo {pages} {58001} (\bibinfo {year}
  {2008})}\BibitemShut {NoStop}%
\bibitem [{\citenamefont {Das}\ \emph {et~al.}(2008)\citenamefont {Das},
  \citenamefont {Horbach},\ and\ \citenamefont {Voigtmann}}]{Das.2008}%
  \BibitemOpen
  \bibfield  {author} {\bibinfo {author} {\bibfnamefont {S.~K.}\ \bibnamefont
  {Das}}, \bibinfo {author} {\bibfnamefont {J.}~\bibnamefont {Horbach}}, \ and\
  \bibinfo {author} {\bibfnamefont {{\relax Th}.}~\bibnamefont {Voigtmann}},\
  }\href@noop {} {\bibfield  {journal} {\bibinfo  {journal} {Phys.~Rev.~B}\
  }\textbf {\bibinfo {volume} {78}},\ \bibinfo {pages} {064208} (\bibinfo
  {year} {2008})}\BibitemShut {NoStop}%
\bibitem [{\citenamefont {Kumagai}\ \emph {et~al.}(2007)\citenamefont
  {Kumagai}, \citenamefont {Nikkuni}, \citenamefont {Hara}, \citenamefont
  {Izumi},\ and\ \citenamefont {Sakai}}]{Kumagai.2007}%
  \BibitemOpen
  \bibfield  {author} {\bibinfo {author} {\bibfnamefont {T.}~\bibnamefont
  {Kumagai}}, \bibinfo {author} {\bibfnamefont {D.}~\bibnamefont {Nikkuni}},
  \bibinfo {author} {\bibfnamefont {S.}~\bibnamefont {Hara}}, \bibinfo {author}
  {\bibfnamefont {S.}~\bibnamefont {Izumi}}, \ and\ \bibinfo {author}
  {\bibfnamefont {S.}~\bibnamefont {Sakai}},\ }\href@noop {} {\bibfield
  {journal} {\bibinfo  {journal} {Mater. Trans.}\ }\textbf {\bibinfo {volume}
  {48}},\ \bibinfo {pages} {1313} (\bibinfo {year} {2007})}\BibitemShut
  {NoStop}%
\bibitem [{\citenamefont {Kordel}\ \emph {et~al.}(2011)\citenamefont {Kordel},
  \citenamefont {Holland-Moritz}, \citenamefont {Yang}, \citenamefont {Peters},
  \citenamefont {Unruh}, \citenamefont {Hansen},\ and\ \citenamefont
  {Meyer}}]{Kordel.2011}%
  \BibitemOpen
  \bibfield  {author} {\bibinfo {author} {\bibfnamefont {T.}~\bibnamefont
  {Kordel}}, \bibinfo {author} {\bibfnamefont {D.}~\bibnamefont
  {Holland-Moritz}}, \bibinfo {author} {\bibfnamefont {F.}~\bibnamefont
  {Yang}}, \bibinfo {author} {\bibfnamefont {J.}~\bibnamefont {Peters}},
  \bibinfo {author} {\bibfnamefont {T.}~\bibnamefont {Unruh}}, \bibinfo
  {author} {\bibfnamefont {T.}~\bibnamefont {Hansen}}, \ and\ \bibinfo {author}
  {\bibfnamefont {A.}~\bibnamefont {Meyer}},\ }\href@noop {} {\bibfield
  {journal} {\bibinfo  {journal} {Phys. Rev. B}\ }\textbf {\bibinfo {volume}
  {83}},\ \bibinfo {pages} {104205} (\bibinfo {year} {2011})}\BibitemShut
  {NoStop}%
\bibitem [{\citenamefont {Teichler}(1996)}]{Teichler.1996}%
  \BibitemOpen
  \bibfield  {author} {\bibinfo {author} {\bibfnamefont {H.}~\bibnamefont
  {Teichler}},\ }\href@noop {} {\bibfield  {journal} {\bibinfo  {journal}
  {Phys. Rev. Lett.}\ }\textbf {\bibinfo {volume} {76}},\ \bibinfo {pages} {62}
  (\bibinfo {year} {1996})}\BibitemShut {NoStop}%
\bibitem [{\citenamefont {Mutiara}\ and\ \citenamefont
  {Teichler}(2001)}]{Mutiara.2001}%
  \BibitemOpen
  \bibfield  {author} {\bibinfo {author} {\bibfnamefont {A.~B.}\ \bibnamefont
  {Mutiara}}\ and\ \bibinfo {author} {\bibfnamefont {H.}~\bibnamefont
  {Teichler}},\ }\href@noop {} {\bibfield  {journal} {\bibinfo  {journal}
  {Phys. Rev. E}\ }\textbf {\bibinfo {volume} {64}},\ \bibinfo {pages} {046133}
  (\bibinfo {year} {2001})}\BibitemShut {NoStop}%
\bibitem [{\citenamefont {Ladadwa}\ and\ \citenamefont
  {Teichler}(2006)}]{Ladadwa.2006}%
  \BibitemOpen
  \bibfield  {author} {\bibinfo {author} {\bibfnamefont {I.}~\bibnamefont
  {Ladadwa}}\ and\ \bibinfo {author} {\bibfnamefont {H.}~\bibnamefont
  {Teichler}},\ }\href@noop {} {\bibfield  {journal} {\bibinfo  {journal}
  {Phys. Rev. E}\ }\textbf {\bibinfo {volume} {73}},\ \bibinfo {pages} {031501}
  (\bibinfo {year} {2006})}\BibitemShut {NoStop}%
\bibitem [{\citenamefont {Teichler}(2011)}]{Teichler.2011}%
  \BibitemOpen
  \bibfield  {author} {\bibinfo {author} {\bibfnamefont {H.}~\bibnamefont
  {Teichler}},\ }\href@noop {} {\bibfield  {journal} {\bibinfo  {journal}
  {Phys. Rev. Lett.}\ }\textbf {\bibinfo {volume} {107}},\ \bibinfo {pages}
  {067801} (\bibinfo {year} {2011})}\BibitemShut {NoStop}%
\bibitem [{\citenamefont {Sciortino}\ and\ \citenamefont
  {Kob}(2001)}]{Sciortino.2001}%
  \BibitemOpen
  \bibfield  {author} {\bibinfo {author} {\bibfnamefont {F.}~\bibnamefont
  {Sciortino}}\ and\ \bibinfo {author} {\bibfnamefont {W.}~\bibnamefont
  {Kob}},\ }\href@noop {} {\bibfield  {journal} {\bibinfo  {journal} {Phys.
  Rev. Lett.}\ }\textbf {\bibinfo {volume} {86}},\ \bibinfo {pages} {648}
  (\bibinfo {year} {2001})}\BibitemShut {NoStop}%
\bibitem [{\citenamefont {Voigtmann}\ and\ \citenamefont
  {Horbach}(2009)}]{Voigtmann.2009}%
  \BibitemOpen
  \bibfield  {author} {\bibinfo {author} {\bibfnamefont {{\relax
  Th}.}~\bibnamefont {Voigtmann}}\ and\ \bibinfo {author} {\bibfnamefont
  {J.}~\bibnamefont {Horbach}},\ }\href@noop {} {\bibfield  {journal} {\bibinfo
   {journal} {Phys. Rev. Lett.}\ }\textbf {\bibinfo {volume} {103}},\ \bibinfo
  {pages} {205901} (\bibinfo {year} {2009})}\BibitemShut {NoStop}%
\bibitem [{\citenamefont {Voigtmann}(2011)}]{Voigtmann.2011b}%
  \BibitemOpen
  \bibfield  {author} {\bibinfo {author} {\bibfnamefont {{\relax
  Th}.}~\bibnamefont {Voigtmann}},\ }\href@noop {} {\bibfield  {journal}
  {\bibinfo  {journal} {EPL}\ }\textbf {\bibinfo {volume} {96}},\ \bibinfo
  {pages} {36006} (\bibinfo {year} {2011})}\BibitemShut {NoStop}%
\bibitem [{\citenamefont {van Beijeren}\ and\ \citenamefont
  {Ernst}(1973{\natexlab{a}})}]{Beijeren.1973}%
  \BibitemOpen
  \bibfield  {author} {\bibinfo {author} {\bibfnamefont {H.~M.}\ \bibnamefont
  {van Beijeren}}\ and\ \bibinfo {author} {\bibfnamefont {M.~H.}\ \bibnamefont
  {Ernst}},\ }\href@noop {} {\bibfield  {journal} {\bibinfo  {journal}
  {Physica}\ }\textbf {\bibinfo {volume} {68}},\ \bibinfo {pages} {437}
  (\bibinfo {year} {1973}{\natexlab{a}})}\BibitemShut {NoStop}%
\bibitem [{\citenamefont {van Beijeren}\ and\ \citenamefont
  {Ernst}(1973{\natexlab{b}})}]{Beijeren.1973a}%
  \BibitemOpen
  \bibfield  {author} {\bibinfo {author} {\bibfnamefont {H.~M.}\ \bibnamefont
  {van Beijeren}}\ and\ \bibinfo {author} {\bibfnamefont {M.~H.}\ \bibnamefont
  {Ernst}},\ }\href@noop {} {\bibfield  {journal} {\bibinfo  {journal}
  {Physica}\ }\textbf {\bibinfo {volume} {70}},\ \bibinfo {pages} {225}
  (\bibinfo {year} {1973}{\natexlab{b}})}\BibitemShut {NoStop}%
\bibitem [{\citenamefont {Meyer}\ \emph {et~al.}(2008)\citenamefont {Meyer},
  \citenamefont {St\"uber}, \citenamefont {Holland-Moritz}, \citenamefont
  {Heinen},\ and\ \citenamefont {Unruh}}]{Meyer.2008}%
  \BibitemOpen
  \bibfield  {author} {\bibinfo {author} {\bibfnamefont {A.}~\bibnamefont
  {Meyer}}, \bibinfo {author} {\bibfnamefont {S.}~\bibnamefont {St\"uber}},
  \bibinfo {author} {\bibfnamefont {D.}~\bibnamefont {Holland-Moritz}},
  \bibinfo {author} {\bibfnamefont {O.}~\bibnamefont {Heinen}}, \ and\ \bibinfo
  {author} {\bibfnamefont {T.}~\bibnamefont {Unruh}},\ }\href@noop {}
  {\bibfield  {journal} {\bibinfo  {journal} {Phys. Rev. B}\ }\textbf {\bibinfo
  {volume} {77}},\ \bibinfo {pages} {092201} (\bibinfo {year}
  {2008})}\BibitemShut {NoStop}%
\bibitem [{\citenamefont {Garandet}\ \emph {et~al.}(2004)\citenamefont
  {Garandet}, \citenamefont {Mathiak}, \citenamefont {Botton}, \citenamefont
  {Lehmann},\ and\ \citenamefont {Griesche}}]{Garandet.2004}%
  \BibitemOpen
  \bibfield  {author} {\bibinfo {author} {\bibfnamefont {J.~P.}\ \bibnamefont
  {Garandet}}, \bibinfo {author} {\bibfnamefont {G.}~\bibnamefont {Mathiak}},
  \bibinfo {author} {\bibfnamefont {V.}~\bibnamefont {Botton}}, \bibinfo
  {author} {\bibfnamefont {P.}~\bibnamefont {Lehmann}}, \ and\ \bibinfo
  {author} {\bibfnamefont {A.}~\bibnamefont {Griesche}},\ }\href@noop {}
  {\bibfield  {journal} {\bibinfo  {journal} {Int. J. Thermophys.}\ }\textbf
  {\bibinfo {volume} {25}},\ \bibinfo {pages} {249} (\bibinfo {year}
  {2004})}\BibitemShut {NoStop}%
\bibitem [{\citenamefont {Huang}\ and\ \citenamefont
  {Chang}(1998)}]{Huang.1998}%
  \BibitemOpen
  \bibfield  {author} {\bibinfo {author} {\bibfnamefont {W.}~\bibnamefont
  {Huang}}\ and\ \bibinfo {author} {\bibfnamefont {Y.~A.}\ \bibnamefont
  {Chang}},\ }\href@noop {} {\bibfield  {journal} {\bibinfo  {journal}
  {Intermetallics}\ }\textbf {\bibinfo {volume} {6}},\ \bibinfo {pages} {487}
  (\bibinfo {year} {1998})}\BibitemShut {NoStop}%
\bibitem [{\citenamefont {Ghosh}(1994)}]{Ghosh.1994}%
  \BibitemOpen
  \bibfield  {author} {\bibinfo {author} {\bibfnamefont {G.}~\bibnamefont
  {Ghosh}},\ }\href@noop {} {\bibfield  {journal} {\bibinfo  {journal} {J.
  Mater. Res.}\ }\textbf {\bibinfo {volume} {9}},\ \bibinfo {pages} {598}
  (\bibinfo {year} {1994})}\BibitemShut {NoStop}%
\bibitem [{\citenamefont {Wang}\ \emph {et~al.}(2011)\citenamefont {Wang},
  \citenamefont {Fang}, \citenamefont {Shang}, \citenamefont {Zhang},
  \citenamefont {Wang}, \citenamefont {Hui}, \citenamefont {Mathaudhu},\ and\
  \citenamefont {Liu}}]{Wang.2011}%
  \BibitemOpen
  \bibfield  {author} {\bibinfo {author} {\bibfnamefont {W.~Y.}\ \bibnamefont
  {Wang}}, \bibinfo {author} {\bibfnamefont {H.~Z.}\ \bibnamefont {Fang}},
  \bibinfo {author} {\bibfnamefont {S.~L.}\ \bibnamefont {Shang}}, \bibinfo
  {author} {\bibfnamefont {H.}~\bibnamefont {Zhang}}, \bibinfo {author}
  {\bibfnamefont {Y.}~\bibnamefont {Wang}}, \bibinfo {author} {\bibfnamefont
  {X.}~\bibnamefont {Hui}}, \bibinfo {author} {\bibfnamefont {S.}~\bibnamefont
  {Mathaudhu}}, \ and\ \bibinfo {author} {\bibfnamefont {Z.~K.}\ \bibnamefont
  {Liu}},\ }\href@noop {} {\bibfield  {journal} {\bibinfo  {journal} {Physica
  B: Cond. Matt.}\ }\textbf {\bibinfo {volume} {406}},\ \bibinfo {pages} {3089}
  (\bibinfo {year} {2011})}\BibitemShut {NoStop}%
\bibitem [{\citenamefont {Sondermann}\ \emph {et~al.}(2014)\citenamefont
  {Sondermann} \emph {et~al.}}]{Sondermann_pre}%
  \BibitemOpen
  \bibfield  {author} {\bibinfo {author} {\bibfnamefont {E.}~\bibnamefont
  {Sondermann}} \emph {et~al.},\ }\href@noop {} {} (\bibinfo {year} {2014}),\
  \bibinfo {note} {{t}o be published}\BibitemShut {NoStop}%
\bibitem [{\citenamefont {Holland-Moritz}\ \emph {et~al.}(2009)\citenamefont
  {Holland-Moritz}, \citenamefont {St\"uber}, \citenamefont {Hartmann},
  \citenamefont {Unruh},\ and\ \citenamefont {Meyer}}]{HollandMoritz.2009}%
  \BibitemOpen
  \bibfield  {author} {\bibinfo {author} {\bibfnamefont {D.}~\bibnamefont
  {Holland-Moritz}}, \bibinfo {author} {\bibfnamefont {S.}~\bibnamefont
  {St\"uber}}, \bibinfo {author} {\bibfnamefont {H.}~\bibnamefont {Hartmann}},
  \bibinfo {author} {\bibfnamefont {T.}~\bibnamefont {Unruh}}, \ and\ \bibinfo
  {author} {\bibfnamefont {A.}~\bibnamefont {Meyer}},\ }\href@noop {}
  {\bibfield  {journal} {\bibinfo  {journal} {J. Phys.: Conf. Ser.}\ }\textbf
  {\bibinfo {volume} {144}},\ \bibinfo {pages} {012119} (\bibinfo {year}
  {2009})}\BibitemShut {NoStop}%
\bibitem [{\citenamefont {Karpe}\ \emph {et~al.}(1995)\citenamefont {Karpe},
  \citenamefont {Krog}, \citenamefont {B\o{}ttiger}, \citenamefont {Chechenin},
  \citenamefont {Somekh},\ and\ \citenamefont {Greer}}]{Karpe.1995}%
  \BibitemOpen
  \bibfield  {author} {\bibinfo {author} {\bibfnamefont {N.}~\bibnamefont
  {Karpe}}, \bibinfo {author} {\bibfnamefont {J.~P.}\ \bibnamefont {Krog}},
  \bibinfo {author} {\bibfnamefont {J.}~\bibnamefont {B\o{}ttiger}}, \bibinfo
  {author} {\bibfnamefont {N.~G.}\ \bibnamefont {Chechenin}}, \bibinfo {author}
  {\bibfnamefont {R.~E.}\ \bibnamefont {Somekh}}, \ and\ \bibinfo {author}
  {\bibfnamefont {A.~L.}\ \bibnamefont {Greer}},\ }\href@noop {} {\bibfield
  {journal} {\bibinfo  {journal} {Acta Metall.\ Mater.}\ }\textbf {\bibinfo
  {volume} {43}},\ \bibinfo {pages} {551} (\bibinfo {year} {1995})}\BibitemShut
  {NoStop}%
\bibitem [{\citenamefont {Pauling}(1947)}]{Pauling.1947}%
  \BibitemOpen
  \bibfield  {author} {\bibinfo {author} {\bibfnamefont {L.}~\bibnamefont
  {Pauling}},\ }\href@noop {} {\bibfield  {journal} {\bibinfo  {journal}
  {J.~Am.~Chem.~Soc.}\ }\textbf {\bibinfo {volume} {69}},\ \bibinfo {pages}
  {542} (\bibinfo {year} {1947})}\BibitemShut {NoStop}%
\bibitem [{\citenamefont {Cordero}\ \emph {et~al.}(2008)\citenamefont
  {Cordero}, \citenamefont {G\'omez}, \citenamefont {Platero-Prats},
  \citenamefont {Rev\'es}, \citenamefont {Echeverr\'\i{}a}, \citenamefont
  {Cremandes}, \citenamefont {Barrag\'aan},\ and\ \citenamefont
  {Alvarez}}]{Cordero.2008}%
  \BibitemOpen
  \bibfield  {author} {\bibinfo {author} {\bibfnamefont {B.}~\bibnamefont
  {Cordero}}, \bibinfo {author} {\bibfnamefont {V.}~\bibnamefont {G\'omez}},
  \bibinfo {author} {\bibfnamefont {A.~E.}\ \bibnamefont {Platero-Prats}},
  \bibinfo {author} {\bibfnamefont {M.}~\bibnamefont {Rev\'es}}, \bibinfo
  {author} {\bibfnamefont {J.}~\bibnamefont {Echeverr\'\i{}a}}, \bibinfo
  {author} {\bibfnamefont {E.}~\bibnamefont {Cremandes}}, \bibinfo {author}
  {\bibfnamefont {F.}~\bibnamefont {Barrag\'aan}}, \ and\ \bibinfo {author}
  {\bibfnamefont {S.}~\bibnamefont {Alvarez}},\ }\href@noop {} {\bibfield
  {journal} {\bibinfo  {journal} {Dalton Trans.}\ ,\ \bibinfo {pages} {2832}}
  (\bibinfo {year} {2008})}\BibitemShut {NoStop}%
\bibitem [{\citenamefont {Plevachuk}\ \emph {et~al.}(2007)\citenamefont
  {Plevachuk}, \citenamefont {Egry}, \citenamefont {Brillo}, \citenamefont
  {Holland-Moritz},\ and\ \citenamefont {Kaban}}]{Plevachuk.2007}%
  \BibitemOpen
  \bibfield  {author} {\bibinfo {author} {\bibfnamefont {Y.}~\bibnamefont
  {Plevachuk}}, \bibinfo {author} {\bibfnamefont {I.}~\bibnamefont {Egry}},
  \bibinfo {author} {\bibfnamefont {J.}~\bibnamefont {Brillo}}, \bibinfo
  {author} {\bibfnamefont {D.}~\bibnamefont {Holland-Moritz}}, \ and\ \bibinfo
  {author} {\bibfnamefont {I.}~\bibnamefont {Kaban}},\ }\href@noop {}
  {\bibfield  {journal} {\bibinfo  {journal} {Int. J. Mater. Res.}\ }\textbf
  {\bibinfo {volume} {98}},\ \bibinfo {pages} {107} (\bibinfo {year}
  {2007})}\BibitemShut {NoStop}%
\bibitem [{\citenamefont {Paradis}\ \emph {et~al.}(2014)\citenamefont
  {Paradis}, \citenamefont {Ishikawa}, \citenamefont {Lee}, \citenamefont
  {Holland-Moritz}, \citenamefont {Brillo}, \citenamefont {Rhim},\ and\
  \citenamefont {Okada}}]{Paradis.2014}%
  \BibitemOpen
  \bibfield  {author} {\bibinfo {author} {\bibfnamefont {P.-F.}\ \bibnamefont
  {Paradis}}, \bibinfo {author} {\bibfnamefont {T.}~\bibnamefont {Ishikawa}},
  \bibinfo {author} {\bibfnamefont {G.-W.}\ \bibnamefont {Lee}}, \bibinfo
  {author} {\bibfnamefont {D.}~\bibnamefont {Holland-Moritz}}, \bibinfo
  {author} {\bibfnamefont {J.}~\bibnamefont {Brillo}}, \bibinfo {author}
  {\bibfnamefont {W.-K.}\ \bibnamefont {Rhim}}, \ and\ \bibinfo {author}
  {\bibfnamefont {J.~T.}\ \bibnamefont {Okada}},\ }\href@noop {} {\bibfield
  {journal} {\bibinfo  {journal} {Mater. Sci. Engin. R}\ }\textbf {\bibinfo
  {volume} {76}},\ \bibinfo {pages} {1} (\bibinfo {year} {2014})}\BibitemShut
  {NoStop}%
\bibitem [{\citenamefont {Davies}\ \emph {et~al.}(2002)\citenamefont {Davies},
  \citenamefont {Dinsdale}, \citenamefont {Gisby}, \citenamefont {Robinson},\
  and\ \citenamefont {Martin}}]{mtbase}%
  \BibitemOpen
  \bibfield  {author} {\bibinfo {author} {\bibfnamefont {R.~H.}\ \bibnamefont
  {Davies}}, \bibinfo {author} {\bibfnamefont {A.~T.}\ \bibnamefont
  {Dinsdale}}, \bibinfo {author} {\bibfnamefont {J.~A.}\ \bibnamefont {Gisby}},
  \bibinfo {author} {\bibfnamefont {J.~A.~J.}\ \bibnamefont {Robinson}}, \ and\
  \bibinfo {author} {\bibfnamefont {S.~M.}\ \bibnamefont {Martin}},\
  }\href@noop {} {\bibfield  {journal} {\bibinfo  {journal} {CALPHAD}\ }\textbf
  {\bibinfo {volume} {26}},\ \bibinfo {pages} {229} (\bibinfo {year} {2002})},\
  \bibinfo {note} {http://resource.npl.co.uk/mtdata/databases.htm}\BibitemShut
  {NoStop}%
\bibitem [{\citenamefont {Cheng}\ and\ \citenamefont {Ma}(2011)}]{Cheng.2011}%
  \BibitemOpen
  \bibfield  {author} {\bibinfo {author} {\bibfnamefont {Y.~Q.}\ \bibnamefont
  {Cheng}}\ and\ \bibinfo {author} {\bibfnamefont {E.}~\bibnamefont {Ma}},\
  }\href@noop {} {\bibfield  {journal} {\bibinfo  {journal} {Prog. Mater.
  Sci.}\ }\textbf {\bibinfo {volume} {56}},\ \bibinfo {pages} {379} (\bibinfo
  {year} {2011})}\BibitemShut {NoStop}%
\bibitem [{\citenamefont {Wang}\ \emph {et~al.}(2014)\citenamefont {Wang},
  \citenamefont {Herlach},\ and\ \citenamefont {Liu}}]{Wang.2014}%
  \BibitemOpen
  \bibfield  {author} {\bibinfo {author} {\bibfnamefont {H.}~\bibnamefont
  {Wang}}, \bibinfo {author} {\bibfnamefont {D.~M.}\ \bibnamefont {Herlach}}, \
  and\ \bibinfo {author} {\bibfnamefont {R.~P.}\ \bibnamefont {Liu}},\
  }\href@noop {} {\bibfield  {journal} {\bibinfo  {journal} {EPL}\ }\textbf
  {\bibinfo {volume} {105}},\ \bibinfo {pages} {36001} (\bibinfo {year}
  {2014})}\BibitemShut {NoStop}%
\bibitem [{\citenamefont {Voigtmann}(2008)}]{Voigtmann.2008d}%
  \BibitemOpen
  \bibfield  {author} {\bibinfo {author} {\bibfnamefont {{\relax
  Th}.}~\bibnamefont {Voigtmann}},\ }\href@noop {} {\bibfield  {journal}
  {\bibinfo  {journal} {Phys. Rev. Lett.}\ }\textbf {\bibinfo {volume} {101}},\
  \bibinfo {pages} {095701} (\bibinfo {year} {2008})}\BibitemShut {NoStop}%
\bibitem [{\citenamefont {Das}\ \emph {et~al.}(2006)\citenamefont {Das},
  \citenamefont {Fisher}, \citenamefont {Sengers}, \citenamefont {Horbach},\
  and\ \citenamefont {Binder}}]{Das.2006}%
  \BibitemOpen
  \bibfield  {author} {\bibinfo {author} {\bibfnamefont {S.~K.}\ \bibnamefont
  {Das}}, \bibinfo {author} {\bibfnamefont {M.~E.}\ \bibnamefont {Fisher}},
  \bibinfo {author} {\bibfnamefont {J.~V.}\ \bibnamefont {Sengers}}, \bibinfo
  {author} {\bibfnamefont {J.}~\bibnamefont {Horbach}}, \ and\ \bibinfo
  {author} {\bibfnamefont {K.}~\bibnamefont {Binder}},\ }\href@noop {}
  {\bibfield  {journal} {\bibinfo  {journal} {Phys. Rev. Lett.}\ }\textbf
  {\bibinfo {volume} {97}},\ \bibinfo {pages} {025702} (\bibinfo {year}
  {2006})}\BibitemShut {NoStop}%
\end{thebibliography}%
\bibliographystyle{apsrev4-1}

\end{document}